\documentclass[aps, pra, nofootinbib, onecolumn, 11pt, tightenlines, notitlepage, superscriptaddress, longbibliography]{revtex4-1}
\usepackage[letterpaper, total={6in, 9in}]{geometry}
\usepackage{amssymb,mathtools}
\usepackage{amsmath,dsfont,physics}
\usepackage{mathrsfs}
\usepackage{graphicx}
\usepackage{dcolumn}
\usepackage[colorlinks=true,allcolors=blue]{hyperref}
\usepackage{siunitx}
\usepackage{verbatim}
\usepackage{natbib}
\usepackage{cleveref}
\usepackage{float}
\usepackage{color,xcolor}
\usepackage[utf8]{inputenc}
\usepackage[english]{babel}
\usepackage{amsthm}
\usepackage{appendix}

\begin{document}

\title{Simulated coherent electron shuttling in silicon quantum dots}

\author{Brandon Buonacorsi}
\affiliation{Institute for Quantum Computing, University of Waterloo, Waterloo, Ontario N2L 3G1, Canada}
\affiliation{Waterloo Institute for Nanotechnology, University of Waterloo, Waterloo, Ontario N2L 3G1, Canada}
\affiliation{Department of Physics and Astronomy, University of Waterloo, Waterloo, Ontario N2L 3G1, Canada}

\author{Benjamin Shaw}
\affiliation{Institute for Quantum Computing, University of Waterloo, Waterloo, Ontario N2L 3G1, Canada}

\author{Jonathan Baugh}
\affiliation{Institute for Quantum Computing, University of Waterloo, Waterloo, Ontario N2L 3G1, Canada}
\affiliation{Waterloo Institute for Nanotechnology, University of Waterloo, Waterloo, Ontario N2L 3G1, Canada}
\affiliation{Department of Chemistry, University of Waterloo, Waterloo, Ontario N2L 3G1, Canada}

\begin{abstract}
Shuttling of single electrons in gate-defined silicon quantum dots is numerically simulated. A minimal gate geometry without explicit tunnel barrier gates is introduced, and used to define a chain of accumulation mode quantum dots, each controlled by a single gate voltage. One-dimensional potentials are derived from a three-dimensional electrostatic model, and used to construct an effective Hamiltonian for efficient simulation. Control pulse sequences are designed by maintaining a fixed adiabaticity, so that different shuttling conditions can be systematically compared. We first use these tools to optimize the device geometry for maximum transport velocity, considering only orbital states and neglecting valley and spin degrees of freedom. Taking realistic geometrical constraints into account, charge shuttling speeds up to $\sim$300 m/s preserve adiabaticity. Coherent spin transport is simulated by including spin-orbit and valley terms in an effective Hamiltonian, shuttling one member of a singlet pair and tracking the entanglement fidelity. With realistic device and material parameters, shuttle speeds in the range $10-100$ m/s with high spin entanglement fidelities are obtained when the tunneling energy exceeds the Zeeman energy. High fidelity also requires the inter-dot valley phase difference to be below a threshold determined by the ratio of tunneling and Zeeman energies, so that spin-valley-orbit mixing is weak. In this regime, we find that the primary source of infidelity is a coherent spin rotation that is correctable, in principle. The results pertain to proposals for large-scale spin qubit processors in isotopically purified silicon that rely on coherent shuttling of spins to rapidly distribute quantum information between computational nodes. 
\end{abstract}

\maketitle

\section{Introduction}
Electron spin qubits in silicon have emerged as a leading platform for scalable quantum information processing in CMOS-like architectures \cite{veldhorst2014addressable,zajac2018resonantly,watson2018programmable,yoneda2018quantum,xue2019benchmarking,huang2019fidelity,sigillito2019coherent}. The small footprint of a gate-defined quantum dot (QD), $\sim$50-100 nm in scale, means that high qubit density is a long-term advantage for scaling, but also brings significant practical challenges. The ability to rapidly transport quantum information over intermediate length scales would mitigate some of these challenges and be a valuable resource from an architecture design perspective. Recent architecture proposals \cite{buonacorsi2018network,li2018crossbar} feature coherent spin shuttling as a primary resource. Shuttling can be used to share entanglement between small neighbouring computational nodes, enabling the 2D surface code to be mapped to a network-of-nodes architecture \cite{buonacorsi2018network}. Separating the scaling problem into intra-node and inter-node operations is advantageous, and creates space for practical wiring interconnects while maintaining a high qubit density compared to state of the art ion trap and superconducting qubit technologies. \\
\indent Coherent transport of quantum information encoded in the electron spin can be realized in several ways. Surface acoustic waves (SAWs) in a piezoelectric material such as GaAs have been used to deterministically transport single charges over several microns \cite{kataoka2009coherent,utko2007single,bertrand2016injection, takada2019sound}. Silicon is not piezoelectric, but a thin ZnO layer was shown to enable SAW-driven charge transport in silicon \cite{buyukkose2013ultrahigh}. One drawback of the SAW approach is that it requires transducers that are large compared to QDs. Another approach is to manipulate the exchange interaction in a linear array of singly-charged QDs. An arbitrary spin state can be transported either via a sequence of SWAP gates \cite{kandel2019coherent} or by an ``all-on" method such as coherent transfer by adiabatic passage (CTAP) \cite{greentree2004coherent,rahman2009atomistic}. This has the advantage of a fixed charge state for all dots, but requires fine-tuned control of tunnel barriers and therefore has a limited resilience to charge and voltage noise. In this paper, we focus on coherent shuttling: electrostatically-driven, sequential tunneling of a single charge/spin through a chain of empty QDs. Coherent spin shuttling was demonstrated in GaAs QD devices \cite{fujita2017coherent,flentje2017coherent}, despite the presence of nuclear-spin induced decoherence. In silicon, shuttling of a single charge across a linear array of nine dots in 50 ns has been reported \cite{mills2019shuttling}. It is anticipated that the weak spin-orbit interaction for electrons in silicon, together with the ability to remove nuclear spins through isotopic purification, could set the stage for maintaining spin coherence over long shuttling distances. Prior theoretical studies have examined the impact of spin-orbit and valley physics on spin transport fidelities \cite{li2017intrinsic,zhao2018coherent}. It was found that the presence of multiple valley states and variation in valley phase can give rise to significant error, although this can be mitigated by operating away from so-called leakage hot spots. \\
\indent In this paper, we connect the shuttling problem to realistic devices, developing tools to optimize both the device geometry and the voltage sequences for shuttling. First, an algorithm for constructing voltage sequences is designed that maintains a constant adiabatic parameter. These constant-adiabaticity control sequences are a useful tool for systematic comparison and optimization, and we use them throughout the paper. The device layout investigated is a simplified MOS geometry in which each accumulation mode QD is formed by a single plunger gate electrode and there are no explicit tunnel barrier gates. Tunneling is controlled both by the voltages on adjacent plunger gates and by the fixed spatial gaps between electrodes. Realistic potentials from a 3D finite element model are mapped to 1D potentials to simulate shuttling along a chain. Charge shuttling in the absence of spin and valley effects is first studied, to test the performance of the adiabatic control sequences and to optimize the device geometry for maximum (adiabatic) speed of transport. The geometry optimization relies on an effective double QD Hamiltonian in which detuning and orbital excitation energies are determined based on the finite element potentials. Subsequently, we extend the effective Hamiltonian to include spin and valley physics, and study the entanglement fidelity after shuttling one member of a spin singlet pair. In the regime that Zeeman energy is smaller than the resonant tunneling energy, we identify a parameter range in which high shuttling fidelities and speeds up to $\sim$80 m/s are possible. The implications of this study on coherent spin transport in $^{28}$Si MOS qubit architectures is discussed.  

\section{Constant-adiabaticity control sequences}
\label{sec:pulseAlgorithm}
For an adiabatic tunneling process, an electron initialized in the orbital ground state, $\ket{\psi_0}$, remains in the ground state at all times. The adiabaticity of the process is quantified by the approximate adiabatic parameter \cite{comparat2009general}
\begin{equation}
    \label{eq:adiabaticCondition}
    \xi(t) = \sum_{m\ne 0}\hbar\left|\frac{\langle{\psi_m(t)}|\frac{d}{dt}|\psi_0(t)\rangle}{E_0(t) - E_m(t)}\right|
\end{equation}
where the index $m$ runs over all excited states, and $E_m(t)$ is the energy of the eigenstate $\ket{\psi_m}$ at time $t$. 
When $\xi(t) \gtrapprox  1$, diabatic transitions to excited orbital states occur with high probability. Conversely, when $\xi(t) \ll 1$, the orbital state retains a large overlap with the ground state.
The condition $\xi(t) \ll 1$ is achieved when the Hamiltonian changes slowly with respect to the frequency corresponding to the ground-excited state gap. \\ 
\indent Tunneling between two QDs is achieved by sweeping the inter-dot detuning $\epsilon = \epsilon_1 - \epsilon_2$, where $\epsilon_i$ corresponds to the orbital ground state energy of the $i^{\rm th}$ QD. In previous theoretical studies \cite{buonacorsi2018network, zhao2018coherent} and experimental demonstrations \cite{mills2019shuttling,fujita2017coherent} of shuttling/tunneling, linear detuning pulses were used. While practically convenient, linear pulses do not maintain constant adiabaticity, and discontinuities in the pulse shape can cause undesired excitations. In order to systematically compare shuttling simulations with different geometrical and voltage parameters, and to optimize the device design for shuttling speed, it is convenient to use pulses that maintain a constant $\xi$. We design such pulses using an algorithm described below. Fidelity of a pulse is defined by the overlap of the final orbital state, in which the electron is located in the target dot, with the ground orbital state in the target dot. The fidelity of an adiabatic pulses can be tuned to an arbitrary value by choice of $\xi$, if only the orbital state is considered (spin and valley physics neglected). \\
\indent Consider a linear chain of $n$ QDs described by the Hamiltonian $H(\vec{V}) = \frac{-\hbar^2}{2m^*}\nabla^2 + v(V_1,...,V_n)$ where $v$ is the electrostatic potential. Here, only the orbital component of the electron wavefunction is considered (spin will be considered in later sections), and we assume there is no ground state degeneracy. $\{V_i\}$ are the voltages applied to the gate electrodes that each define an individual accumulation-mode QD and tune the energy levels $\epsilon_i$. The set of these voltage parameters is vectorized as $\vec{V}$. We wish to find a pulse sequence $\vec{V}(t)$ that shuttles the electron through the $n$-dot chain while keeping $\xi$ fixed. In later sections, we will use an effective Hamiltonian expressed directly in terms of the dot potentials $\epsilon_i$. In that case, the vector of dot potentials $\vec{\epsilon}(t)$ is input into the the algorithm as the set of control variables. The algorithm is presented below for a double QD system, but readily generalizes to an $n$-dot chain.   
\begin{enumerate}
    \item Choose voltage configurations $\{\vec{V}(A), \vec{V}(B), \vec{V}(C)\}$ at three time points (A, B, C) that the Hamiltonian should pass through during the shuttling process.
    \begin{enumerate}
        \item $\vec{V}(A)$ tunes $H$ so that the electron is fully localized in QD $\#1$ ($\epsilon < 0$).
        \item $\vec{V}(B)$ tunes $H$ so that the electron resonantly tunnels between the two QDs ($\epsilon = 0$).
        \item $\vec{V}(C)$ tunes $H$ so that the electron is fully localized in QD $\#2$ ($\epsilon > 0$).
    \end{enumerate}
    \item Select a sufficiently large number, $N$, of voltage configurations interpolated between $\vec{V}(A)$, $\vec{V}(B)$ and $\vec{V}(C)$, and choose a desired adiabatic parameter $\xi'$.
    \item For each interpolated voltage configuration $\vec{V}(i)$:
    \begin{enumerate}
        \item Solve for the eigenstates of the Hamiltonians $H(\vec{V}(i))$ and $H(\vec{V}(i) + \delta\vec{V}(i))$ where $\delta\vec{V}(i)$ is a small voltage difference.
        \item Use the calculated eigenstates and $\delta\vec{V}(i)$ to approximate $\frac{d}{d\vec{V}(i)}|\psi_0(\vec{V}(i))\rangle$.
        \item Find $\frac{d\vec{V}(i)}{dt}$ such that $\frac{d\vec{V}(i)}{dt}\frac{d}{d\vec{V}(i)}|\psi_0(\vec{V}(i))\rangle$ when used in Equation \ref{eq:adiabaticCondition} gives $\xi = \xi'$.
    \end{enumerate}
    \item Let $\vec{V}(t_i)$ correspond to voltage configuration $\vec{V}(i)$ at time $t_i$. Set the initial condition as $\vec{V}(t_0 = 0) = \vec{V}(A)$. Then convert each voltage configuration index $V(i)$ to $V(t_i)$ by $t_i = t_{i-1} + \frac{dt}{d\vec{V}(i)}(\vec{V}(i+1) - \vec{V}(i))$.
\end{enumerate}
The algorithm does not assume a fixed pulse duration, but converges to a certain length based on the chosen value of $\xi$. Convergence requires selecting a sufficiently large number of interpolation points in step 2 ($N$ is deemed sufficiently large when the final pulse does not vary with increasing $N$). The relationship between the applied voltages $\vec{V}$ and the electrostatic potential is evaluated using a self-consistent 3D Poisson solver based on the chosen device geometry (this is not required when using the effective Hamiltonians of sections~\ref{sec:optimalGeom} and ~\ref{sec:effSimulations} expressed directly in terms of the dot potentials $\epsilon_i$). A large set of gate voltage configurations are simulated in order to provide a `library' of potential landscapes to be used in the algorithm. The discrete set of potentials are interpolated to provide a quasi-continuous distribution (step 2). We approximate the true potentials by ignoring the effect of the single electron charge and solving the Poisson equation in the limit of zero charge density. While quantitatively approximate, this allows us to qualitatively study shuttling dynamics while avoiding the technical difficulty of maintaining a fixed charge in a Schr\"{o}dinger-Poisson solver. In Appendix \ref{app:schrodingerVSpoisson}, the effect of an electron charge on a double QD potential is calculated, showing that at resonant tunneling, reduction of the tunnel barrier height is the main effect. This can be compensated for by suitable adjustment of the gate geometry and pulse design. In the effective Hamiltonian simulations of section~\ref{sec:optimalGeom}, we use the Schr\"{o}dinger-Poisson method to determine orbital energy spacings and to determine the tunnel coupling as a function of double QD geometry. \\
\indent Longer QD chains are treated by adding more voltage configurations at step 1 ($2n-1$ configurations for shuttling through $n$ dots). For example, shuttling to a third dot is realized by including configurations $\{\vec{V}(D), \vec{V}(E)\}$. It is assumed that there is no ground state degeneracy during shuttling, as this causes Eq.~\ref{eq:adiabaticCondition} to diverge and the algorithm to fail. Shuttling pulses can also be found for an electron in the $k^{\rm th}$ excited state by substituting $|\psi_k\rangle$ for $|\psi_0\rangle$ in Eq.~\ref{eq:adiabaticCondition}, assuming the orbital relaxation rate is slow compared to shuttling. Our approach for designing adiabatic control pulses is valid for any Hamiltonian of the form $H = H_0 + H_c(u_\alpha, u_\beta,\dots)$ where $H_0$ is static and $H_c$ is a time-varying term with control parameters $\{u_\alpha, u_\beta, \dots\}$. However, if $H$ is either complex or contains oscillatory terms, then evolution under the pulse may not adiabatic, as Eq.~\ref{eq:adiabaticCondition} does not guarantee adiabaticity for Hamiltonians of that form \cite{comparat2009general}.

\section{Charge shuttling}
\label{sec:coherentShuttling}
\begin{figure}
    \centering
    \includegraphics[width = \textwidth]{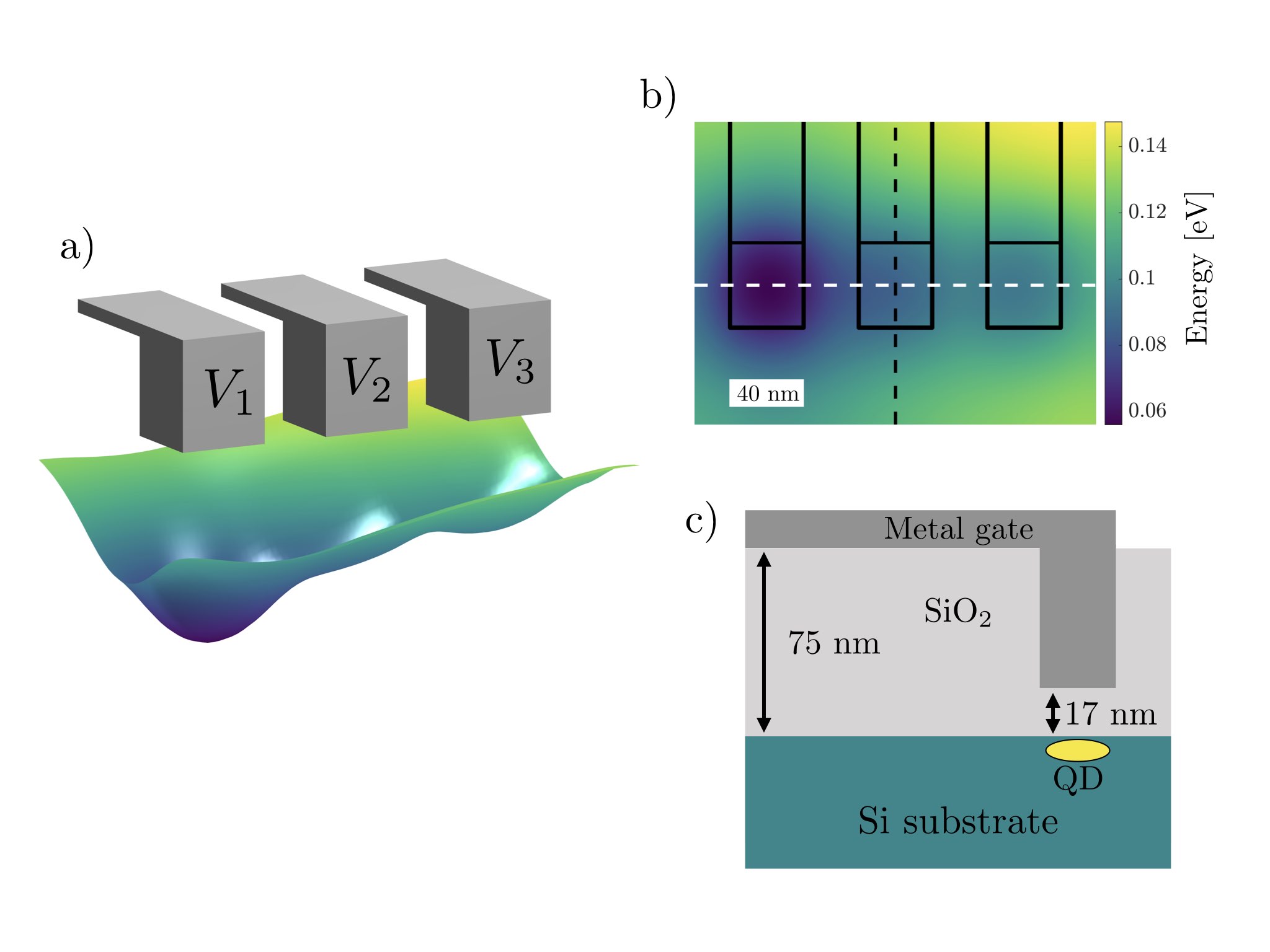}
    \caption{Schematic of a triple linear quantum dot chain using a `via' gate geometry with no explicit tunnel barrier gates. a) 3D render of the gate geometry with a plot of a simulated electrostatic potential obtained with a self-consistent Poisson calculation . b) A 2D top view of the potential with plunger gates outlined. The potential is a 2D slice taken 1 nm below the Si/SiO$_2$ interface. Darker color indicates a more attractive potential for electrons. The white horizontal dashed line indicates the 1D potential slice used in the shuttling simulations. c) Side-view of a plunger gate taken along the black line in (b) showing the vertical plunger gate design. The yellow oval indicates electron accumulation in a quantum dot.}
    \label{fig:QDchainSchematic}
\end{figure}

\begin{figure}[h]
    \centering
    \includegraphics[width = \textwidth,trim = {0 8cm 0 8cm},clip]{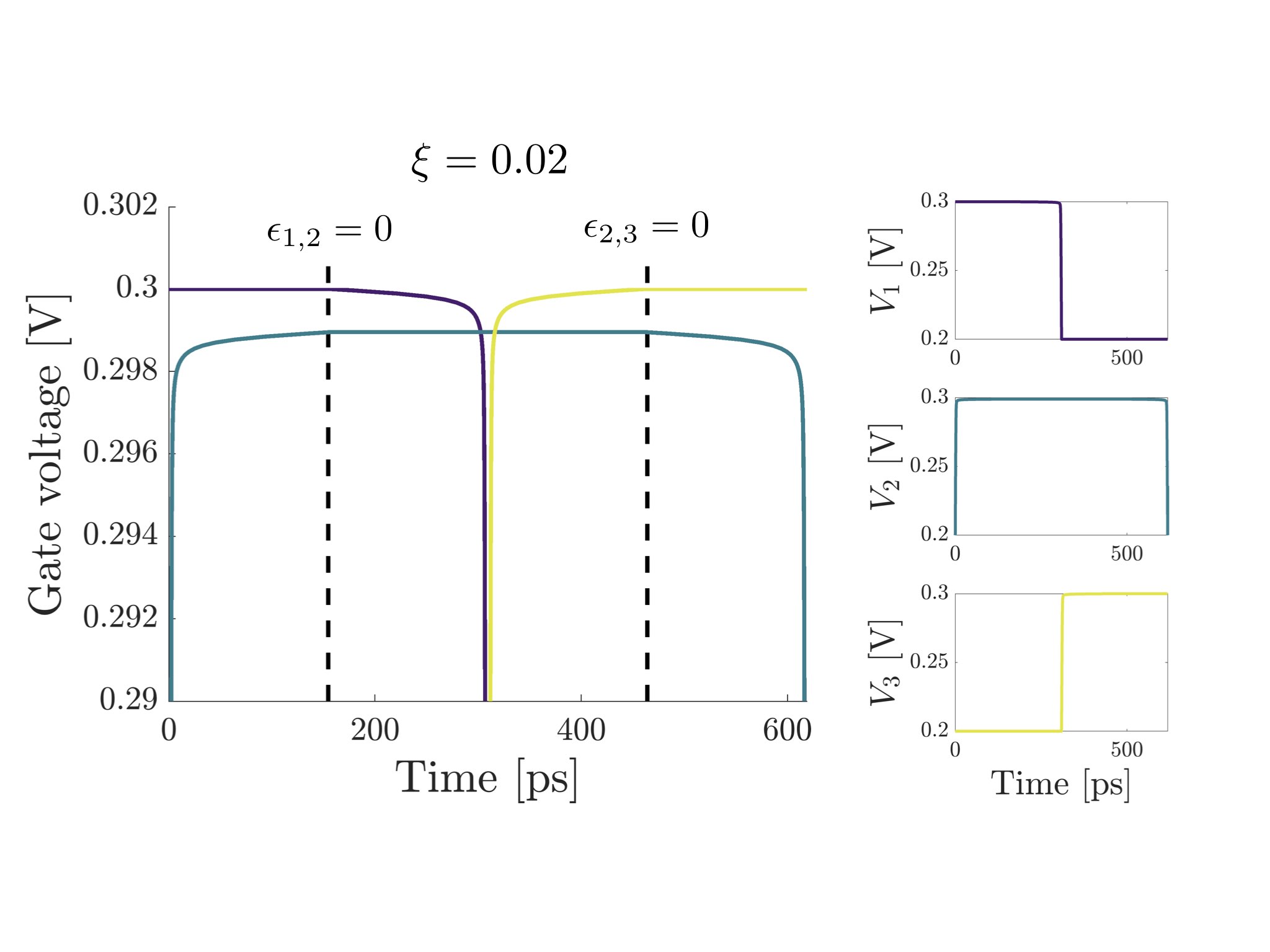}
    \caption{A constant-adiabaticity shuttling pulse calculated for the linear triple dot system, with $\xi = 0.02$. The electron is initially localized in dot 1 and then shuttled through dots 2 and 3 by sweeping the three plunger gates. When $V_1 \approx V_2$ or $V_2 \approx V_3$, the detuning between neighbouring dots is $\epsilon_{i,j} = \epsilon_i - \epsilon_j \approx 0$. The right inset figures show the pulse shapes over the full voltage range, while the main (left) panel is an enlarged view showing the smooth nature of the pulse shape near the upper corners.}
    \label{fig:vPulse}
\end{figure}

This section investigates the performance of adiabatic pulses by simulating electron shuttling along a triple QD linear chain, considering only single-valley orbital states and neglecting both spin and valley physics. This pertains to the physical case of charge shuttling, in which the metric of interest is the fidelity of remaining in the orbital ground state. In the single-valley case the ground state is unique; in the presence of valley physics, it is a ground-state manifold. While the simulations presented in this section neglect multiple valley states, we confirmed that the results are equivalent to having equal valley splittings and zero valley phase difference between adjacent dots, with valley degrees of freedom traced out at the end of the calculation (see Appendix~\ref{app:singleValleyApprox}). The purpose of this section is also to introduce the device geometry of interest, the use of constant-adiabaticity pulses, and the optimization of device geometry with respect to the speed and fidelity of charge shuttling. \\
\indent Each accumulation-mode QD is defined by a single plunger gate, and there are no explicit gates to control tunnelling barriers \cite{buonacorsi2018network,ramirez2018few}.
Tunnelling is controlled both by the applied gate voltages and the fixed geometric gap separating adjacent gates. The full device structure, including the metal gates and the Si/SiO$_2$ heterostructure, is simulated using a self-consistent 3D Poisson solver in nextnano++ \cite{birner2007nextnano} (refer to Appendix \ref{app:schrodingerVSpoisson} for details).\\
\indent Figure \ref{fig:QDchainSchematic}a shows a 3D view of a triple QD model and a 2D slice of a simulated potential landscape taken 1 nm below the Si/SiO$_2$ interface. The corresponding plunger gate voltages were $V_1 = 0.3$ V and $V_2 = V_3 = 0.2$ V. Figure \ref{fig:QDchainSchematic}b shows a 2D top view of the potential landscape with an outline of the plunger gates superimposed. The plunger gate heads are 40 nm $\times$ 40 nm, and the edge to edge separation between them is 30 nm. Figure \ref{fig:QDchainSchematic}c shows a side view of the device structure taken along the black dotted line in Figure \ref{fig:QDchainSchematic}b. This view highlights the plunger gate's vertical design in which electrons only accumulate below the thinner oxide section, which is 17 nm thick in this model. 
The pulse control parameters are the plunger gate voltages $\{V_1, V_2, V_3\}$. Approximately 1000 potentials were calculated using plunger gate voltage configurations ranging from $[0.2, 0.3]$ V in steps of 0.01 V for each gate. Potentials at voltage configurations in between these points are obtained by linear interpolation. The potential term in the Hamiltonian is $v(V_1, V_2, V_3)$ where $v$ is a 1D slice of the potential landscape taken along the white dashed line in Figure \ref{fig:QDchainSchematic}c, 1 nm below the Si/SiO$_2$ interface. A 1D potential is used here to reduce computational resources, but 2D or 3D potentials could be used, in principle. \\
\indent A constant-adiabaticity pulse for electron shuttling using $\xi = 0.02$ and a voltage range of $[0.2, 0.3]$ V is plotted in Figure \ref{fig:vPulse}. The left panel is an enlarged view that shows the smooth nature of the pulses near the corners. At time $T = 0$, $V_1 = 0.3$ V and $V_2 = V_3 = 0.2$ V which localizes the electron in dot 1. At $T \approx 155$ ns, $\epsilon = 0$ ($V_1 \approx V_2$) and the electron resonantly tunnels between dots 1 and 2. $V_1$ is swept to $V_1 = 0.2$ V at $T \approx 315$ ns which fully localizes the electron in dot 2. A similar process is carried out to shuttle the electron from dot 2 to dot 3. When the detuning $|\epsilon| \gg 0$, gate voltages can be swept quickly without harming adiabicity since the ground-excited state energy gap is large. When $\epsilon \approx 0$ (at $T \approx 155$ ns and $T \approx 465$ ns), the gap is small and the voltages must be swept slowly to maintain adiabaticity. The 1D potentials calculated with the Poisson solver naturally take into account cross-capacitances. This manifests as the zero detuning point ($\epsilon = 0$) occurring at $V_1 > V_2$ rather than $V_1 = V_2$, for example. The dot-to-dot shuttle duration in this example is about 325 ps.

\begin{figure}
    \centering
    \includegraphics[width = \textwidth]{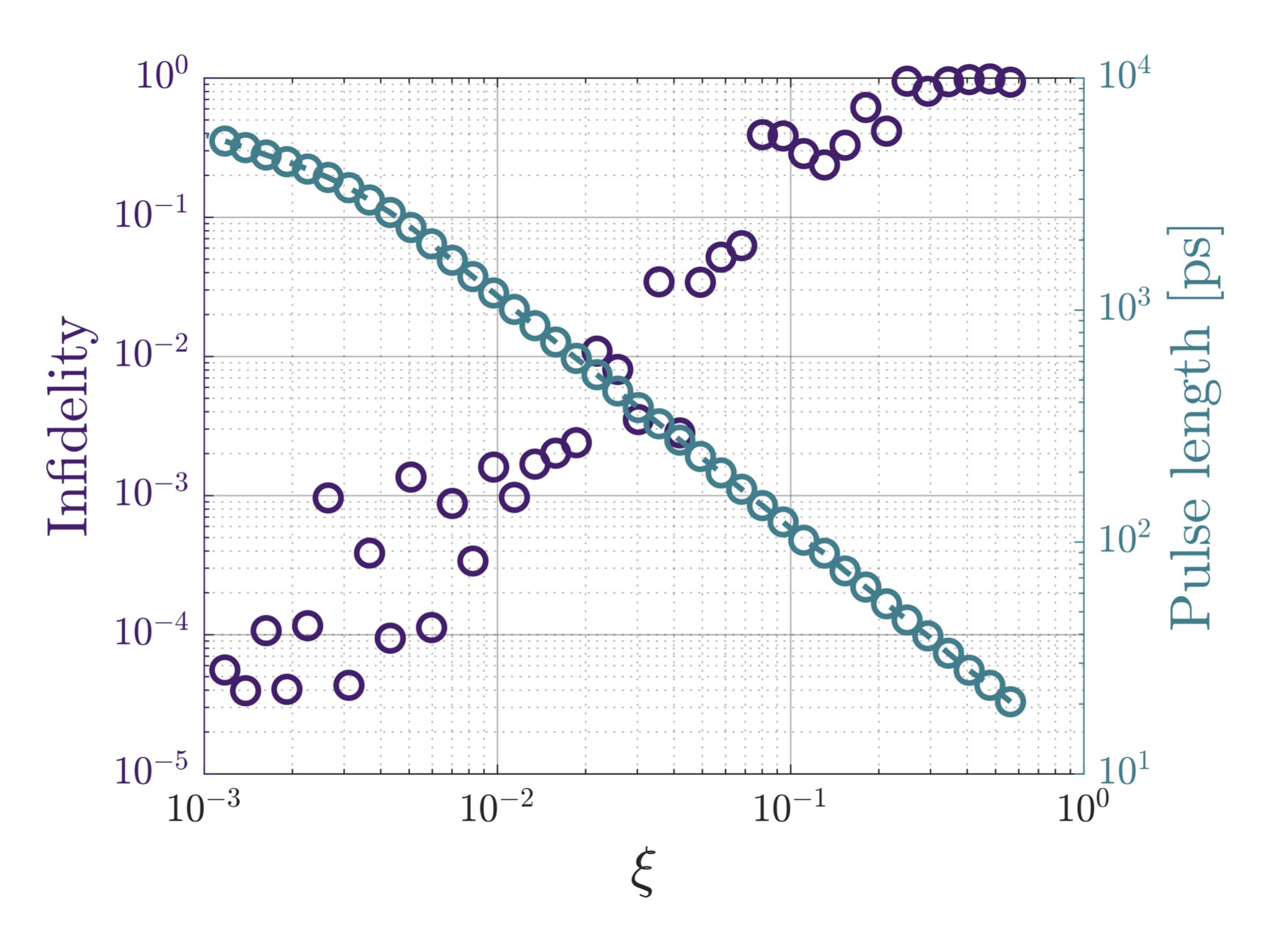}
    \caption{Relationship between the adiabatic parameter $\xi$, final orbital state infidelity $1 - F(T)$, and pulse length $T$. Pulses with arbitrarily high fidelity can be found by reducing $\xi$ at the cost of increased pulse length.}
    \label{fig:fidVSadiab}
\end{figure}

We now examine the fidelity of shuttling using the constant-adiabaticity pulses described previously. The electron is initialized in the orbital ground state of the potential $v(\vec{V}(0))$. State evolution is calculated by solving the time-dependent Schr\"{o}dinger equation (TDSE). Numerically, the TDSE is solved using the split-operator approach \cite{dion2014program} with a time step $\Delta t = $ \num{5E-16} s. The instantaneous fidelity of the orbital state with the ground state is defined as $F(t) = |\langle\psi_0(t)|\psi_{\rm sim}(t)\rangle|^2$, where $|\psi_0(t)\rangle$ is the ground state for the potential $v(\vec{V}(t))$, and $|\psi_{\rm sim}(t)\rangle$ is the simulated orbital state of the shuttled electron. The quality of a pulse of length $T$ is defined as the final orbital state fidelity $F(T)$. We note that decoherence in the charge basis is neglected in these simulations. Figure \ref{fig:fidVSadiab} summarizes the trade-off between final orbital state infidelity $1 - F(T)$ and pulse duration $T$ as the adiabatic parameter is varied (see Appendix \ref{app:simResults} for an explicit comparison between an adiabatic and non-adiabatic shuttling).  
    
Apart from $\xi$, the resonant tunnel coupling $t_c$ between two neighboring QDs determines the pulse length. The slowest parts of the pulse occur at the $\epsilon = 0$ anti-crossings where the energy spacing between the ground and first excited orbital state is $2|t_c|$. In the device geometry considered here, there are no gates to directly tune the tunnel barriers between dots. Instead, $t_c$ is determined by the geometry of the gate electrodes and the inter-electrode gaps, as well as the applied gate voltages. For the device geometry in Figure \ref{fig:QDchainSchematic}, $t_c \approx 25$ $\mu$eV which gives sub-nanosecond shuttling pulses with orbital state fidelities $>99\%$. We used a similar geometry and $t_c$ values in a previous study \cite{buonacorsi2018network} with linear pulses. The present results show a threshold time of $\approx$ 325 ps per dot-to-dot shuttling step for a final orbital state fidelity $> 99$\% - a factor of 5 improvement in speed over linear pulses. While superior to linear pulses, constant-adiabaticity pulses are not time optimal, and we expect that faster high-fidelity pulses could be designed by allowing $\xi$ to vary and using optimal control methods. 

\subsection{Device geometry optimization}\label{sec:optimalGeom}
\begin{figure}[h]
    \centering
    \includegraphics[width = \textwidth]{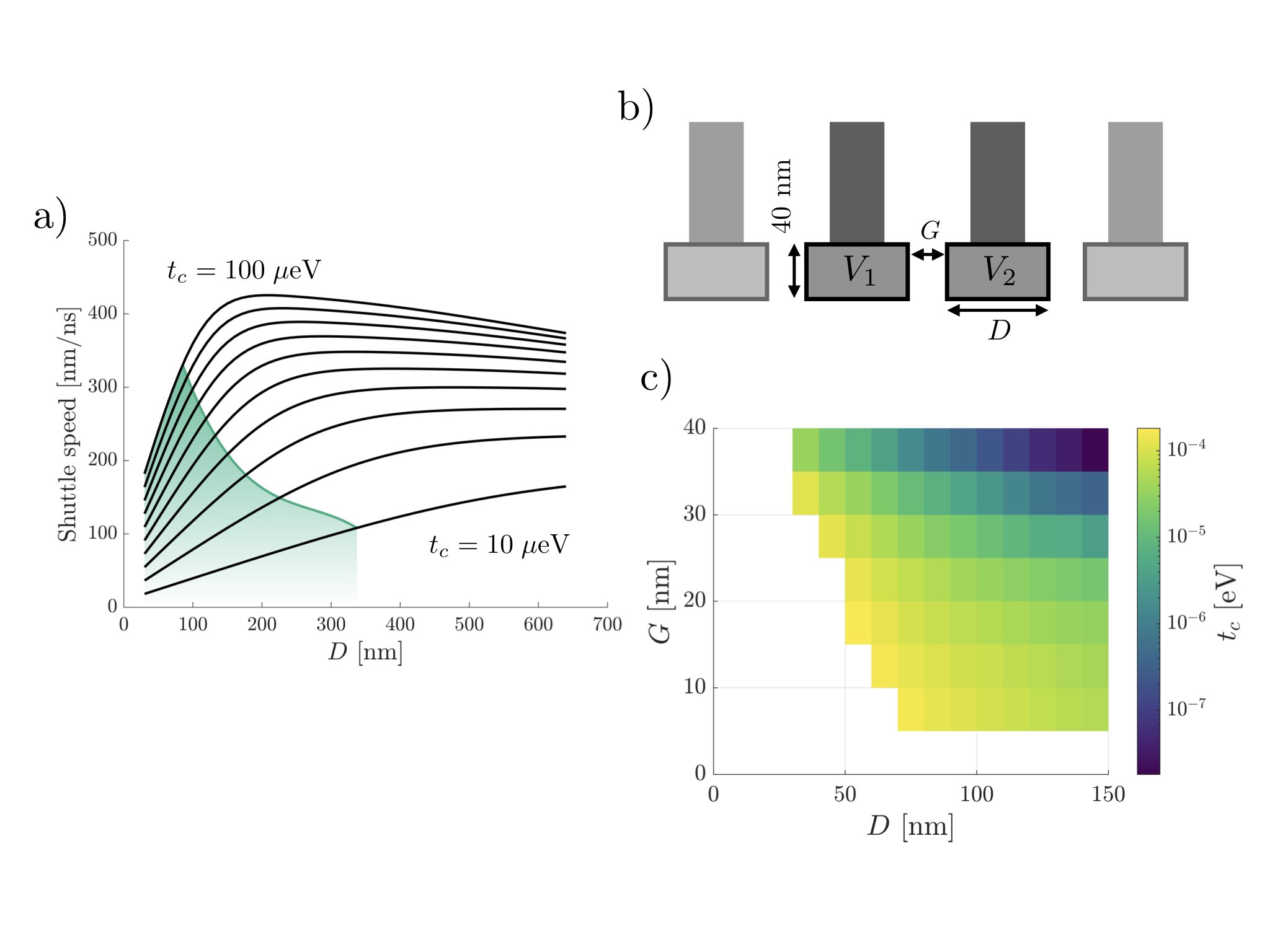}
    \caption{Optimizing the gate electrode geometry for fast shuttling. a) Shuttling velocity $(G + D)/T$ determined by finding a constant-adiabaticity pulse of duration $T$ for given values of $D$ and $t_c$. The ten curves corresponds to values of $t_c$ ranging from 10 to 100 $\mu$eV in steps of 10 $\mu$eV. The shaded (green) region corresponds to the range of ($D$, $t_c$) values achievable in the geometry of (b) with $G \geq 10$ nm. Smaller gap values are considered impractical for realistic device fabrication. b) Top-down schematic view of the four-electrode model used to calculate $t_c$ as a function of geometrical parameters $G$ and $D$. The two central gates form the double QD used to model shuttling, while the outer gates are set to fixed potentials to make the double dot potential realistic. The dot length $D$ and inter-electrode gap $G$ are varied uniformly for all four gates. c) Dependence of $t_c$ on $D$ and $G$ resulting from a 3D Schr\"{o}dinger-Poisson calculation of the four-electrode model, with a single electron occupying the central double dot.}
    \label{fig:optimalGeometry}
\end{figure}

In this section, we use simulations of the constant-adiabaticity control pulses to optimize the gate electrode design of Figure \ref{fig:QDchainSchematic} for maximum shuttling velocity. Stretching out the QDs in the direction of transport increases the distance travelled per shuttle, however, this also reduces the QD orbital energy spacing and requires slower pulses to maintain adiabaticity. To investigate this trade-off, we find shuttling pulses for double QDs with varying plunger gate length $D$ and gate separation $G$ and quantify the shuttling speed. An effective, approximate Hamiltonian describing the orbital dynamics of the shuttled electron is used. The Hamiltonian for the double QD is
\begin{equation}
\label{eq:effHamiltonian_excitedOrb}
    H = \begin{bmatrix} \epsilon_L & 0 & t_c & t_c \\
    0 & \epsilon_L + \Delta E_L & t_c & t_c \\
    t_c & t_c & \epsilon_R & 0 \\
    t_c & t_c & 0 & \epsilon_R + \Delta E_R
    \end{bmatrix}
\end{equation}
where $\epsilon_k$ is the ground state energy of each dot, $\Delta E_d$ is the ground to first excited orbital splitting of each dot, $t_c$ is the resonant tunnel coupling, and $d=L,R$ refers to the left and right dots, respectively. Here $t_c$ is treated as an independent parameter and not a function of the dot size; we will discuss the dependence of $t_c$ on the dot geometry below. The orbital spacing $\Delta E=\Delta E_L=\Delta E_R$ is determined as a function of dot size $D$ using a self-consistent Schr\"{o}dinger-Poisson solver. The potential landscape of a central QD tuned to one-electron occupancy is calculated, embedded within a triple dot so that the outer wings of the central dot's potential landscape are realistic. The outer gate electrodes are set to a fixed potential that tune the outer dots to zero electron occupancy. Details of the calculation are given in Appendix \ref{app:powerLawFit}. The dot potentials $\epsilon_{L,R}$ are used as the control parameters. \\
\indent For various QD lengths $D$, constant-adiabaticity pulses are found with $\xi = 0.005$ and inserting $\Delta E(D)$ into the Hamiltonian above. Once the pulse length $T$ is known for given parameters $D$ and $t_c$, the shuttling velocity is $(G + D)/T$, where $G$ is the inter-electrode gap set to 30 nm. Figure \ref{fig:optimalGeometry}a summarizes the relationship between shuttling velocity and dot size $D$ for different tunnel couplings ranging from $t_c = 10~\mu$eV to 100~$\mu$eV. The shuttling speed initially increases with $D$, but then saturates at a maximum value and gradually decreases thereafter. The initial positive slope is due to a greater distance covered per shuttle step, but as $D$ further increases, the effect of reduced orbital energy spacing dominates, increasing the time $T$ needed to maintain adiabaticity. As expected, the shuttle velocity is a monotonically increasing function of $t_c$.\\
\indent Above, we took $t_c$ as a chosen parameter, however in practice $t_c$ will be determined by a combination of the geometrical parameters ($G$, $D$) and the applied voltages. To get a sense of the range of practical tunnel coupling values, $t_c$ for a double QD was calculated using a 3D Schr\"{o}dinger-Poisson solver over a range of ($G$, $D$) values. Figure \ref{fig:optimalGeometry}b shows the modelled four dot geometry, where the outer gate voltages were fixed at $-0.1$ V and the central gates set to $V_1 = V_2 = V$, with $V$ tuned such that a single electron occupies the symmetric inner double QD potential. The splitting of the lowest two eigenenergies determines $t_c$. The results are plotted in figure \ref{fig:optimalGeometry}c, where $t_c$ decreases monotonically as both $D$ and $G$ increase, with a higher sensitivity to variation in $G$. Assuming a practical fabrication limit of $G =$ 10 nm, achievable electron velocities in Figure \ref{fig:optimalGeometry}a are restricted to the green shaded region. The solid green line bounding the shaded region corresponds to $G = 10$ nm, whereas $G > 10$ nm for the rest of the shaded region. The highest practical shuttle velocities for this device geometry, $\sim 0.3~\mu$m ns$^{-1}$, occur for $D \approx 100$ nm and $G \approx 10$ nm. On the other hand, to reduce the number of shuttle steps, $D$ can be extended to $\approx 300$ nm at the cost of reducing the velocity by a factor $\sim$3. These results demonstrate that a simplified gate geometry (one electrode per QD) can be optimized for single electron shuttling, without the need for additional gates to tune tunnel couplings - effectively reducing the required number of electrodes by 2 for a linear shuttling array.  \\

\section {Spin and valley effects}
\label{sec:effSimulations}
Thus far, we have only considered the electronic orbital state dynamics in a single-valley setting and ignored spin. For quantum information processors based on QDs in silicon, shuttling of electron spin qubits, especially one member of an entangled pair, could be a critically important resource \cite{buonacorsi2018network,li2018crossbar}. To examine this possibility, an effective Hamiltonian model that accounts for spin and valley degrees of freedom is used to study the limits of coherent single spin transport by shuttling. We use the entanglement fidelity of a two-spin state to gauge the fidelity of the process, however, the second spin is considered to be static and never physically close to the shuttled spin. \\
\subsection{Valley-orbit Hamiltonian}
\begin{figure}[h]
    \centering
    \includegraphics[width = \textwidth]{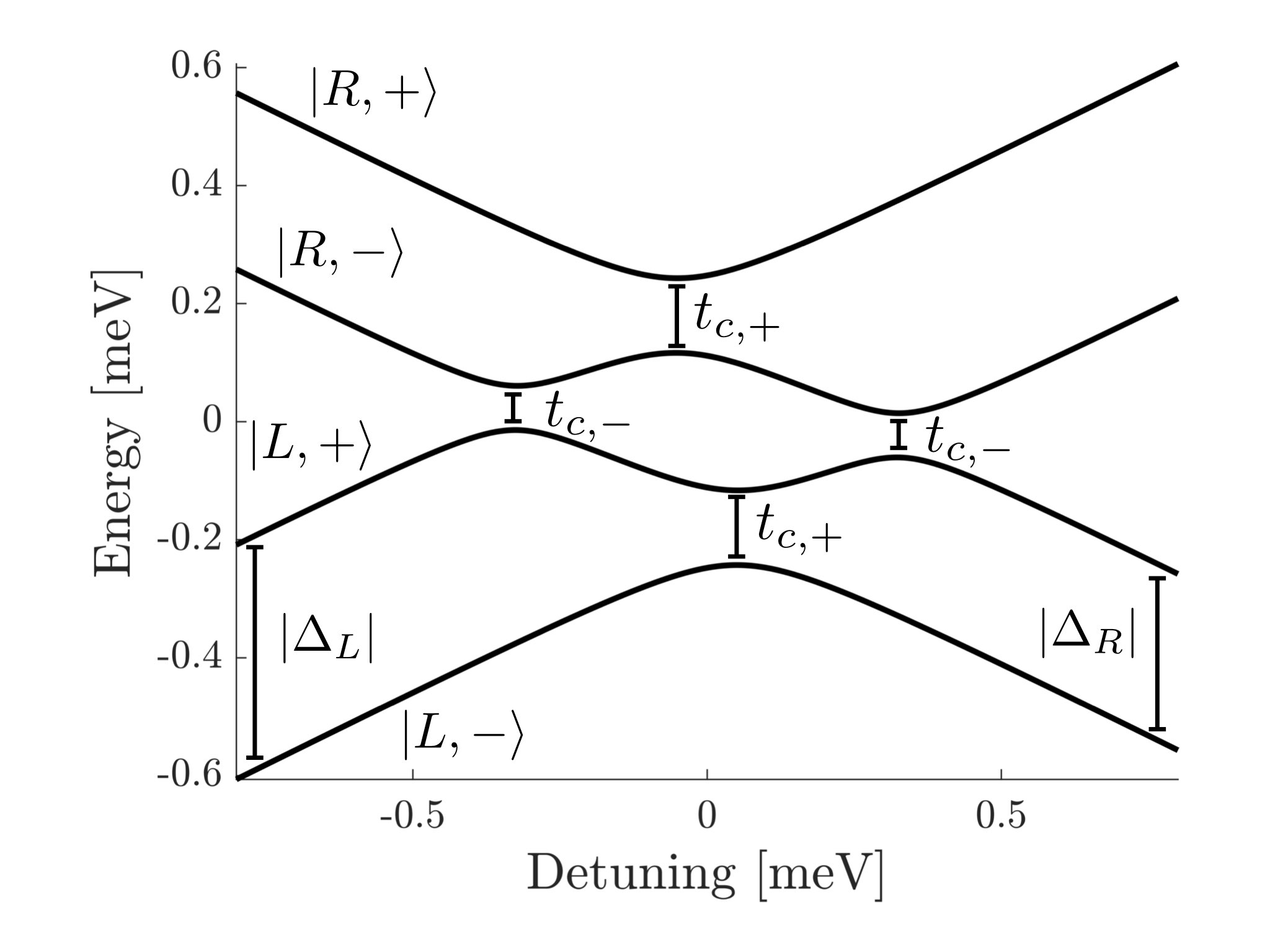}
    \caption{Valley-orbit energy spectrum for a double quantum dot with single electron occupation. The Hamiltonian parameters are $|\Delta_L| = 200$ $\mu$eV, $|\Delta_R| = 150$ $\mu$eV, and $\delta\phi = \pi/3$. The four eigenstates when $\epsilon \ll 0$ are labelled on the left side. The intra-valley and inter-valley tunnel couplings are labelled $t_{c,+}$ and $t_{c,-}$, respectively.}
    \label{fig:valleyOrbit}
\end{figure}

Bulk silicon has six-fold degenerate conduction band minima referred to as valleys. In a Si/SiO$_2$ hetero-structure, strong confinement along the vertical ($\hat{z}$) direction and strain at the Si/SiO$_2$ interfaces raises the energy of the four in-plane valleys, leaving a 2-fold degeneracy of the out-of-plane valley states $\ket{z}$ and $\ket{\tilde{z}}$. The sharp change in potential at the interface couples $\ket{z}$ and $\ket{\tilde{z}}$, lifting the degeneracy and giving two valley eigenstates $\ket{\pm} = \frac{1}{\sqrt{2}}(\ket{z} \pm e^{i\phi}\ket{\tilde{z}})$. The eigenstates $\ket{\pm}$ are separated in energy by the valley splitting $\Delta = |\Delta|e^{i\phi}$, where $\phi$ is the phase of the electron's Bloch wave function \cite{tagliaferri2018impact,culcer2010quantum}. Disorder at the Si/SiO$_2$ interface causes $|\Delta|$ and $\phi$ to vary randomly between QDs \cite{culcer2010interface,gamble2016valley}. \\
\indent The small but non-zero spin-orbit coupling in silicon's conduction band mixes spin and valley eigenstates and is a source of spin decoherence for shuttled electrons \cite{yang2013spin, ferdous2018valley}. Satisfying the adiabatic condition Eq.~\ref{eq:adiabaticCondition} is not alone sufficient for maintaining coherence, since for example, the Hamiltonian is only real if the valley phase difference between adjacent QDs is zero, and adiabatic evolution is not guaranteed if the Hamiltonian is not real \cite{comparat2009general}. \\
\indent The valley phase difference between two neighboring QDs, $\delta\phi = \phi_1 - \phi_2$, can strongly affect how fast the electron can be adiabatically shuttled, as we discuss below. For one electron occupying a double QD, there are two charge orbital configurations $\ket{L}$ and $\ket{R}$ corresponding to the electron occupying the left and right QDs, respectively. These orbitals are coupled with tunneling strength $t_c$. In the single electron valley-orbit Hamiltonian, there are four anti-crossings formed from two types of inter-dot tunnel couplings. The \textit{intra-valley} tunnel coupling $t_{c,+} = \frac{t_c}{2}(1 + e^{-i\delta\phi})$ allows tunneling events between QD orbitals with the same valley eigenstate ($\ket{L,\pm}$ and $\ket{R,\pm}$), whereas the \textit{inter-valley} tunnel coupling $t_{c,-} = \frac{t_c}{2}(1 - e^{-i\delta\phi})$ couples opposite valley eigenstates ($\ket{L,\pm}$ and $\ket{R,\mp}$) (see Appendix \ref{app:effSVOHam} for more detail) \cite{zhao2018coherent}. Figure \ref{fig:valleyOrbit} shows a valley-orbit energy diagram for a silicon double QD with the $t_{c,-}$ and $t_{c,+}$ anti-crossings labelled. The two $t_{c,+}$ and two $t_{c,-}$ anti-crossings occur at energies $\epsilon = \pm(|\Delta_L| - |\Delta_R|)$ and $\epsilon = \pm(|\Delta_L| + |\Delta_R|)$, respectively. Sweeping the inter-dot detuning $\epsilon = \epsilon_1 - \epsilon_2$ adiabatically through any of these four anti-crossings moves an electron from one QD to the other.
As discussed in Section \ref{sec:coherentShuttling}, the tunnel coupling $t_{c}$ for an electron in the ground state limits the adiabatic shuttling speed. The couplings $t_{c,+}$ and $t_{c,-}$ are dependent on $\delta\phi$. The highest shuttling speed is obtained when $\delta\phi = 0$ and $|t_{c,+}| = t_c$. With increasing $\delta\phi$, $|t_{c,+}|$ decreases, requiring longer constant-adiabaticity pulses. For $\delta\phi = \pi$, $|t_{c,+}| = 0$ and intra-valley tunneling is completely suppressed. At the $t_{c,-}$ anti-crossing, the opposite occurs; inter-valley tunneling cannot occur for $\delta\phi = 0$, whereas for $0 < \delta\phi < \pi$, $|t_{c,-}|$ is finite and yields an anti-crossing that mixes the $\ket{L,\pm}$ and $\ket{R,\mp}$ valley-orbit states. When $\delta\phi = \pi$, $|t_{c,-}| = t_c$ and the inter-valley gap is completely opened.\\
\subsection{Spin transport}
\begin{figure}[h]
    \centering
    \includegraphics[width = \textwidth]{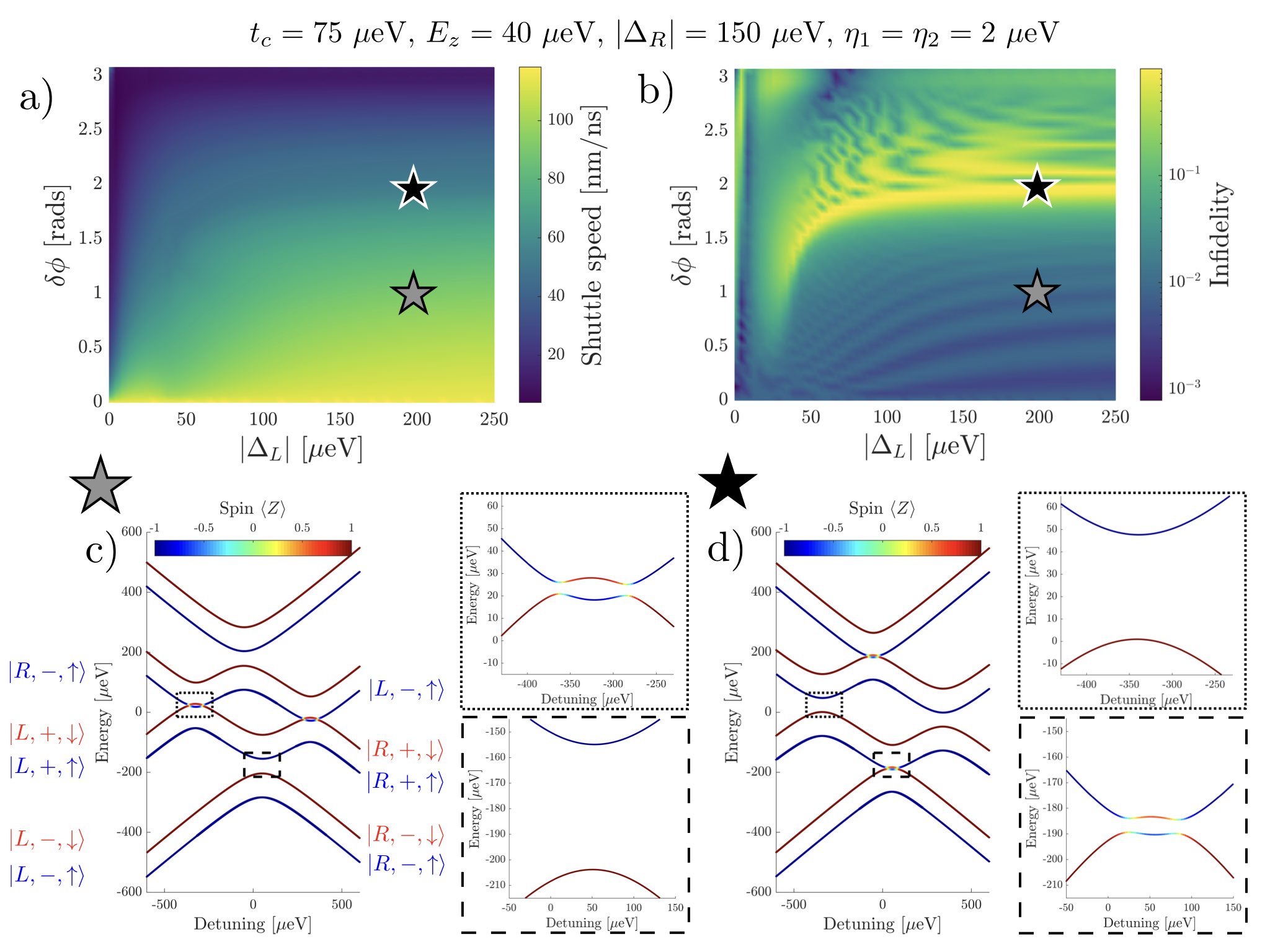}
    \caption{Shuttling one member of a singlet pair, for $t_c > E_z$.  For all panels, the fixed parameters are: $\xi=0.005$, $t_c$ = 75 $\mu$eV, $E_z$ = 40 $\mu$eV, $|\Delta_R|$ = 150 $\mu$eV, $\eta_1$ = $\eta_2$ = 2 $\mu$eV. a) Variation of shuttle speed (colour scale) with the left QD valley splitting $|\Delta_L|$ and the inter-dot valley phase difference $\delta\phi$. These speeds are based on finding constant adiabaticity pulses with $\xi=0.005$. b) The fidelity of maintaining the spin singlet state versus $|\Delta_L|$ and $\delta\phi$. Infidelity is plotted in colour scale, defined as $1-|{\rm Tr}[\rho(T)(I_4\otimes\ket{S}\bra{S})]|^2$, where $\rho(T)$ is the density matrix of the post-shuttle state. High (low) fidelity is indicated by dark blue (yellow). c), d) Energy spectra of the Hamiltonian in Eq.~\ref{eq:effHamiltonian} versus detuning $\epsilon_L - \epsilon_R$. $|\Delta_L| = 200$ $\mu$eV in both panels, $\delta\phi = 1$ rad and $\delta\phi = 2$ rad for (c) and (d), respectively. Colour indicates the spin state, with red (blue) corresponding to spin down (up). Energy levels are labelled by the corresponding eigenstates on the left when the detuning $\ll 0$ and on the right when the detuning $\gg 0$. Enlarged views near the $t_{c,+}$ (dashed square) and $t_{c,-}$ (dotted square) anti-crossings illustrate how SVO mixing varies with $\delta\phi$.}
    \label{fig:effResultstcGTEz}
\end{figure}

\begin{figure}[h]
    \centering
    \includegraphics[width = 0.9\textwidth]{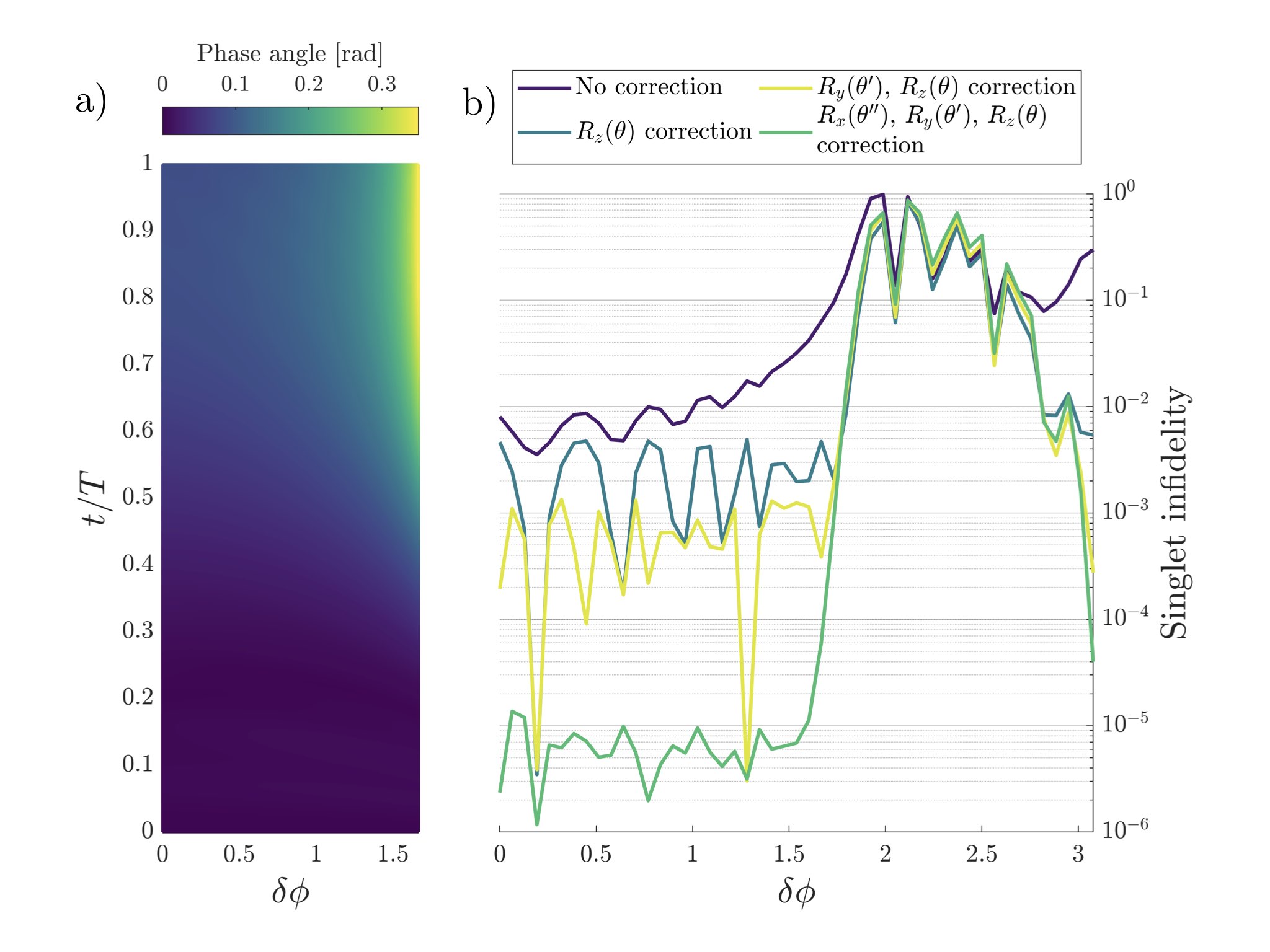}
    \caption{Error due single-spin rotation during shuttling. For both panels, $\xi = 0.005$, $t_c$ = 75 $\mu$eV $> E_z$ = 40 $\mu$eV, $|\Delta_L|$ = 200 $\mu$eV, $|\Delta_R|$ = 150 $\mu$eV, and $\eta_1$ = $\eta_2$ = 2 $\mu$eV. a) Phase ($\sigma_z$) rotation of the shuttled spin in the regime $\delta\phi \in [0,1.67]$ rad. b) Effect of corrective rotations on the infidelity of the post-shuttle state with respect to the singlet, as a function of $\delta\phi$. (Purple) no corrective rotations are applied; (blue) $R_z(\theta)$ correction applied; (yellow) $R_z(\theta)$ and $R_y(\theta')$ corrections applied; (green) $R_z(\theta)$, $R_y(\theta')$, and $R_x(\theta'')$ corrections applied.}
    \label{fig:corrSingFid}
\end{figure}

For electron shuttling to be useful in a spin-based quantum information processing device, it must retain the coherence of the spin state. The silicon material system is promising in this respect, since the conduction band spin-orbit coupling is weak compared to that in III-V materials, and  nuclear magnetism can be greatly suppressed by isotopic purification. We now incorporate spin along with orbital and valley degrees of freedom into a double QD effective Hamiltonian model. Only the two lowest valleys are considered, and orbital excited states are neglected, since they are high in energy compared to typical valley splittings. Consider the preparation of a two-electron spin singlet state $\ket{S} = \frac{1}{\sqrt{2}}(\ket{\uparrow\downarrow} - \ket{\downarrow\uparrow})$, with one spin stationary outside of the double dot (e.g. in a third adjacent dot), and the other electron shuttled from left to right within the double dot system. The spin transport fidelity is quantified by the overlap of the post-shuttle spin state with the singlet.\\ 
\indent The effective double QD Hamiltonian is
\begin{multline}
    \label{eq:effHamiltonian}
    H = \left[\sum_{d = L,R}(\epsilon_d k_d \otimes \tau_0 \otimes s^1_0) + t_c k_x \otimes \tau_0 \otimes s^1_0 + \sum_{d=L,R}(\Delta_d k_d \otimes \tau_+ \otimes s^1_0 + h.c.) \right. \\
    \left. \vphantom{\sum_{d = L,R}} + E_z k_0 \otimes \tau_0 \otimes s^1_z + \eta_1 k_z \otimes \tau_0 \otimes s^1_x + \eta_2 k_y \otimes \tau_0 \otimes s^1_y \right] \otimes s_0 + E_z k_0 \otimes \tau_0 \otimes s_0^1 \otimes s_z^2
\end{multline}
where the bracketed terms act on the shuttled electron \cite{zhao2018coherent} and the term outside the bracket acts on the static electron. Two-level operators that act on the orbital, valley and spin subspaces are denoted by $k$, $\tau$ and $s^i$, respectively. In terms of a dummy two-level operator $A$, the operators appearing in Eq.~\ref{eq:effHamiltonian} are defined as $A_{L} = \frac{1}{2}(I + \sigma_z$), $A_{R} = \frac{1}{2}(I - \sigma_z$), $A_0 = I$, $A_x= \sigma_x$, $A_y= \sigma_y$, $A_z= \sigma_z$, and $A_{\pm} = \frac{1}{2}(\sigma_x \pm i \sigma_y)$, where $\sigma_j$ are the Pauli matrices and $I$ is the identity matrix. The 16 basis states are defined by the binary values of the variables $\{d,\nu,s^1, s^2\}$, where $d = L, R$ (left and right orbital ground states), $\nu = z, \tilde{z}$ (valley states), $s^i = \uparrow, \downarrow$ (spin eigenstates of the shuttled [$i = 1$] and stationary [$i = 2$] electrons). Note that a different valley basis is used here ($\{\ket{z}, \ket{\tilde{z}}\}$) than in the previous section; this is because the Hamiltonian in Eq. \ref{eq:effHamiltonian} takes on a more compact form (see Appendix \ref{app:effSVOHam}). The ground state energy for the $n^{\rm th}$ QD is $\epsilon_n$, and $t_c$ is the inter-dot resonant tunnel coupling. The valley splitting of the $d^{\rm th}$ QD is $\Delta_d = |\Delta_d|e^{i\phi_d}$, where $\phi_d$ is the valley phase. $E_z$ is the Zeeman energy due to the static magnetic field. $\eta_1$ and $\eta_2$ describe the spin-orbit interaction with $\eta_1 = \bra{L,\nu,\uparrow}H_{SO}\ket{L,\nu,\downarrow}$ and $\eta_2 = \bra{L,\nu,\uparrow}H_{SO}\ket{R,\nu,\downarrow}$, where $H_{SO}$ is the  spin-orbit Hamiltonian including Rashba and Dresselhaus terms \cite{bychkov1984oscillatory,dresselhaus1955spin}.\\
\indent An electron shuttling from $L$ to $R$ is simulated using the parameters $\{\epsilon_L, \epsilon_R\}$ to define an adiabatic pulse with $\xi = 0.005$. The detuning $\epsilon = \epsilon_L - \epsilon_R$ is swept from -600 $\mu$eV to +600 $\mu$eV. The initial state is $\ket{\psi(0)} = \frac{1}{\sqrt{2}}\ket{\psi^{VO}_0(0)}\otimes(\ket{\uparrow\downarrow} - \ket{\downarrow\uparrow})$ where $\ket{\psi^{VO}_0(0)}$ is the ground state of the initial valley-orbit Hamiltonian. The state evolution is calculated by a discretized time-dependent Schr\"{o}dinger equation. For each simulation, we calculate both the shuttle speed and the fidelity of the final spin state with respect to the singlet. An effective speed is based on the electron travel of 60 nm per shuttle, and corresponding duration of the adiabatic pulse $T$. The fidelity of maintaining the singlet state is $|{\rm Tr}[\rho(T)(I_4\otimes\ket{S}\bra{S})]|^2$, where $\rho(T)$ is the density matrix describing the post-shuttle state and $I_4$ is the $4\times4$ identity matrix. \\
\indent Figure \ref{fig:effResultstcGTEz} shows the dependence of shuttle speed and spin fidelity on varying the valley splitting in the left dot, $|\Delta_L|$, and the valley phase difference, $\delta\phi$. The fixed Hamiltonian parameters are $t_c$ = 75 $\mu$eV $> E_z$ = 40 $\mu$eV, $|\Delta_R|$ = 150 $\mu$eV, and $\eta_1$ = $\eta_2$ = 2 $\mu$eV. The chosen spin-orbit strength $\eta_{1,2}$ is about an order of magnitude larger than an experimentally reported value \cite{hao2014electron}. $|\Delta|_L$ is varied from from 0.1 - 250 $\mu$eV, and $\delta\phi$ from $[0,\pi)$ rad ($\delta\phi = \pi$ is excluded because the ground state is degenerate at that point). The range $\delta\phi = (\pi, 2\pi]$ would produce a mirror image. 
Since $t_c > E_z$, the lowest energy states $\ket{L,-,\uparrow}$ and $\ket{L,-,\downarrow}$ form a ground state manifold. The energy gap with respect to this manifold, set by $|\Delta_L|$ and $t_{c,+}$, determines the speed of the constant-adiabaticity pulse. This is evident in Figure \ref{fig:effResultstcGTEz}a, where the shuttle speed decreases as $\delta\phi$ increases, due to the closing of the $|t_{c,+}|$ gap. The gap closing can be seen in panels (c) and (d) of Figure \ref{fig:effResultstcGTEz}, which show the energy spectra at $\Delta_L = 200$ $\mu$eV for $\delta\phi = 1$ rad and $\delta\phi = 2$ rad, respectively. Spin-valley-orbit (SVO) mixing is evident in the dashed box of panel (d) when $\delta\phi=2$ rad. \\
\indent The valley splitting in the left dot, $|\Delta_L|$, has no significant impact on the shuttle speed as long as $|\Delta_L| > |t_{c,+}|$ ($|\Delta_R|$ is fixed at 150 $\mu$eV in these simulations). However, when $|\Delta_L| \leq |t_{c,+}|$, $|\Delta_L|$ represents the lowest excitation energy and therefore determines the shuttle speed. The crossover point, where $|\Delta_L| = |t_{c,+}|$, moves to smaller $|\Delta_L|$ values as $\delta \phi$ increases. This is the reason why in Figure \ref{fig:effResultstcGTEz}a, for a fixed $\delta \phi$ value such as 1 rad, the shuttle speed increases with $|\Delta_L|$. \\
\indent Figure \ref{fig:effResultstcGTEz}b plots the infidelity (with respect to the singlet) of the post-shuttle spin state versus $\delta\phi$ and $|\Delta_L|$. Fidelities $>$95\% are obtained when $|\Delta_L| > E_z$ and $\delta\phi$ is below about 2 rad. This corresponds to energy spectra qualitatively similar to Figure \ref{fig:effResultstcGTEz}c, where the ground state manifold does not overlap with the lowest excited state. During the adiabatic pulse, the spin-entangled electron occupies only the ground state spin doublet and maintains coherence. When $\delta\phi \approx 2$ rad, the $t_{c,+}$ gap begins to close and SVO mixing occurs, rapidly degrading the state fidelity. The threshold value of $\delta\phi$ for this crossover is given approximately by $E_z \approx |t_{c,+}| = \frac{t_c}{2}|1 + e^{i\delta\phi_{th}}|$, when $|\Delta_L| > E_z$. For the parameters $t_c = 75$ $\mu$eV and $E_z = 40$ $\mu$eV used here, $\delta\phi_{th} \approx 2$ rad. For valley phase differences well above this threshold, the state fidelity improves modestly; this is due to a suppression of SVO mixing near the $t_{c,+}$ anti-crossing as $\delta\phi \rightarrow \pi$. The coupling of spin-orbit eigenstates through the $\eta_1$ term is governed by $t_{c,+}$ and $t_{c,-}$ (see Appendix \ref{app:effSVOHam}). If either $t_{c,+}$ or $t_{c,-}$ equals 0, $\eta_1$ does not cause SVO mixing near the corresponding anti-crossing. In the valley eigenbasis, there are two distinct $\eta_2$ couplings: $\eta_{2,\pm} = \frac{\eta_2}{2}(1 \pm e^{-i\delta\phi})$. As with $t_{c,\pm}$, $\eta_{2,+}$ ($\eta_{2,-}$) couple intra-valley (inter-valley) spin-orbit states. The $\eta_{2,+}$ term mixes states at the $t_{c,+}$ anti-crossing, but approaches zero as $\delta\phi \rightarrow \pi$. \\
\indent In the high fidelity shuttling regime, where $|\Delta_L| > E_z$ and $\delta\phi < \delta\phi_{\rm th}$, the infidelity is primarily caused by a precession of the shuttled electron's spin state about an effective axis due to the presence of the spin-orbit $\eta_1$ and $\eta_2$ terms in addition to the Zeeman term. In other words, the singlet state is not an eigenstate of the spin Hamiltonian when the spin-orbit coupling terms are non-zero. The dominant error is a phase rotation of the singlet into the $\ket{T_0} = \frac{1}{\sqrt{2}}(\ket{\uparrow\downarrow} + \ket{\downarrow\uparrow})$ triplet state. Figure \ref{fig:corrSingFid}a plots the phase rotation angle with $|\Delta_L| =$ 200 $\mu$eV and $\delta\phi \in [0,1.7]$ rad.  The normalized shuttling time $t/T$ is given on the $y$-axis, where $T$ is the total pulse length for the constant adiabaticity shuttling pulses ($\xi = 0.005$). $T$ increases with $\delta\phi$, causing the spin to accumulate a larger phase error at larger $\delta\phi$ values. \\
\indent Figure \ref{fig:corrSingFid}b shows that in addition to phase rotation, the finite spin-orbit terms lead to small rotations about $\sigma_x$ and $\sigma_y$ as well. Fidelity with the singlet state significantly improves as corrective rotations $R_n(\theta) = \exp(-i\theta\sigma_n/2)$, where $n=\{x,y,z\}$, are applied to the shuttled electron spin. The correction angles required for the $R_z(\theta)$, $R_y(\theta')$ and $R_x(\theta'')$ rotations are found by calculating the overlap between the final state and the $\ket{T_0}$, $\frac{1}{\sqrt{2}}(\ket{\uparrow\uparrow}+\ket{\downarrow\downarrow})$, and $\frac{1}{\sqrt{2}}(\ket{\uparrow\uparrow}-\ket{\downarrow\downarrow})$ states, respectively. The trace in Figure \ref{fig:corrSingFid}b with no corrective rotations (purple) is a line cut along $\delta\phi$ from Figure \ref{fig:effResultstcGTEz}b, with $|\Delta_L| =$ 200 $\mu$eV. When $\delta\phi < \delta\phi_{\rm th}$, corrective rotations significantly improve the singlet fidelity. Above $\delta\phi_{\rm th}$, the SVO mixing during shuttling produces a spin state with purity $< 1$ upon tracing out the orbital and valley degrees of freedom. As $\delta\phi$ approaches $\pi$, however, it can be seen that the corrective rotations again improve fidelity due to suppression of SVO mixing near the $t_{c,+}$ anti-crossing. \\
\indent The $R_z(\theta)$ corrections remove the dominant spin rotation error, resulting in $\approx 99.5\%$ singlet fidelity below $\delta\phi_{\rm th}$. Additional $R_y(\theta')$ and $R_x(\theta'')$ corrections further improve fidelity by nearly three orders of magnitude, giving a singlet infidelity $\approx 10^{-5}$. The remaining error after applying all three corrective rotations is due to weak SVO mixing from the $\eta_{1,2}$ Hamiltonian terms. When all three corrective pulses are applied, small variation of the adiabatic parameter $\xi$ does not affect the singlet fidelity, indicating that the state evolution in these simulations is well inside the adiabatic regime. If any of the single spin corrections are not applied, however, slower pulses (smaller $\xi$) will make the fidelity worse, as more single-spin rotation error accumulates. With all corrections applied, pulses with smaller $\xi$ (more adiabatic) slightly enlarge the high-fidelity region of $\delta \phi$ by reducing SVO mixing near the gap-closing threshold $\delta\phi_{\rm th}$.\\
\indent In the regime $|\Delta_L| < E_z$ in Figure \ref{fig:effResultstcGTEz}b, the corresponding energy spectra are more complex. The states labeled (at large negative detuning) $\ket{L,-\downarrow}$ and $\ket{L,+,\uparrow}$ overlap near zero detuning, irrespective of $\delta\phi$. This explains the funnel-shaped, low-fidelity feature at low $|\Delta_L|$ and $\delta\phi$ values. One naturally asks what happens in the other regime, $t_c < E_z$? There, a similar state overlap occurs for almost all $\delta\phi$ and $|\Delta_L|$ values, so the $t_c < E_z$ regime is unfavourable for spin shuttling. Detailed simulation results in the $t_c < E_z$ regime are given in Appendix \ref{app:effResultsContinued}. \\

\section {Discussion}
The key results of this study can be summarized as follows. In section~\ref{sec:optimalGeom}, it was shown that single electron shuttling is possible using a simplified device geometry in which there is a single gate electrode per dot. Such a geometry would be highly economical for large-scale devices, reducing the required number of electrodes by two, and simplifying the applied voltage sequences. In the single valley case, adiabatic transport is achieved at speeds up to 0.3 $\mu$m/ns. Transport speed is mainly determined by the resonant tunneling energy $t_c$, which can reach the 100 $\mu$eV scale in our simplified device geometry with practical fabrication constraints considered. In section~\ref{sec:effSimulations}, we studied the entanglement fidelity of a shuttled electron spin in the presence of valley states $\ket{z}$ and $\ket{\tilde{z}}$ and a small, but finite, spin-orbit coupling. It was found that the $t_c > E_z$ regime is favourable for high spin fidelity, but only for interdot valley phase differences $\delta \phi$ below a threshold value ($\approx 2$ rad for the parameters used in our simulation). Below this threshold, SVO mixing is weak, and the primary effect of the spin-orbit coupling is to generate systematic single-spin rotations that can, in principle, be corrected. With such corrections applied, very high fidelities $\sim 0.9999$ are recovered, compared to $0.995$ with phase correction only, and $>0.95$ with no corrections. For $\delta\phi \ll \delta \phi_{th}$ and $|\Delta| \gg E_z$, average speed and fidelity (without single-qubit corrections) are estimated as 80 nm/ns and 0.99, respectively. Note that for spin-orbit couplings set to zero, there is no SVO mixing in any of the parameter space, which would result in near-perfect spin fidelities. For $\delta \phi$ values at or above the threshold, strong SVO mixing significantly harms the spin fidelity. The regime of high Zeeman field, $t_c < E_z$, has strong SVO mixing at nearly all values of $\delta \phi$, and is therefore unfavourable for spin shuttling. Thus, variability of the valley phase and remaining in the $t_c > E_z$ regime are two key experimental concerns. It should be noted that our simulations only pertain to the case of well-separated electrons, and do not apply to the initial steps of separating a singlet originating in a single dot. In that regime, a two-electron simulation including electron-electron interactions is necessary, and is left for future work.\\
\indent What are the implications of these results for coherent spin transport, a key resource for large-scale quantum computer architectures in silicon? For entanglement distribution in a network architecture, Nickerson et al showed that a raw fidelity $\sim$0.9 is sufficient, since even one round of entanglement distillation can increase the fidelity to fault tolerant levels \cite{nickerson2013topological}. Consider a chain of 16 dots, with 15 shuttle events to transport an electron from dot 1 to dot 16. Each dot-to-dot shuttle requires a fidelity of $\sim$0.993 for the whole process to be above the $0.9$ threshold. In the regime of $\delta\phi < 1.5$ rad and $|\Delta| > 50~\mu$eV of figure \ref{fig:effResultstcGTEz}, the singlet fidelity is $>0.99$ on average. Applying corrective phase rotations $R_z(\theta)$ increases the fidelity to $\sim 0.995$, which is sufficient for a 16-dot process with fidelity $>0.9$. These values correspond to a spin-orbit coupling strength $\eta_{1,2} = 2$ $\mu$eV, an order of magnitude larger than what has been reported in silicon \cite{hao2014electron}. Simulations with a weaker spin-orbit coupling $\eta_1 = \eta_2 = 0.4$ $\mu$eV are shown in Appendix \ref{app:effResultsContinuedSmallSO}. In the same regime $\delta\phi < 1.5$ rad and $|\Delta| > 50~\mu$eV, the singlet fidelity is $>0.999$ without any corrective rotations, which is sufficient for the 16-dot process. The timescale of this 16-dot shuttle, $\sim$12 ns, is shorter than the fastest single-qubit gates that have been implemented for silicon spin qubits \cite{yoneda2018quantum}. Intermediate scale shuttling, therefore, is not necessarily a speed bottleneck for a processor. Indeed, a 9-dot shuttle in 50 ns has already been demonstrated experimentally \cite{mills2019shuttling}. Moreover, the same experiment showed it is possible to shuttle multiple electrons in parallel (separated by a few dots), so that entanglement distillation would not require a doubling of shuttling times, but would require additional ancilla dots and measurements. On the other hand, the scenarios discussed above assume all dots lie within the parameter space for high fidelity shuttling; a single outlier with sufficiently large valley phase difference or small valley splitting would spoil the scheme. It remains to be seen experimentally whether material quality and device processing can yield sufficient control over these parameters. \\
\section{Conclusions}
In summary, the first half of this work showed how to construct constant-adiabaticity control pulses for shuttling single electrons along a 1D chain of QDs. By keeping the adiabatic parameter constant while varying geometric device parameters, for example, we can compare shuttling under different conditions, and optimize for shuttle speed or fidelity. Our method of simulation connects the 3D device model to an effective Hamiltonian in 1D. The second half modeled coherent spin transport by including spin-orbit and valley terms in an effective Hamiltonian, and shuttling one member of a spin-entangled pair. We found that a high-fidelity process requires $t_c > E_z$, $\delta\phi < \delta \phi_{th}$, and $|\Delta| > E_z$. The threshold value $\delta \phi_{th}$ is a function of the ratio $t_c/E_z$. Shuttle speeds up to 0.3 $\mu$m/ns were obtained in the single-valley case, and up to 80 nm/ns in the two-valley case with spin-orbit coupling present. Our results indicate that disorder-induced variation in the valley phase, if sufficiently large, is a primary obstacle to high-fidelity spin shuttling in $^{28}$Si. Future work includes designing faster pulses (constant-adiabaticity is not time-optimal), shuttling in larger arrays, and including charge noise \cite{krzywda2020adiabatic} and charge dephasing effects. Developing 2D simulations would enable simulating shuttling through a T-junction, a likely feature of realistic device architectures. \\

\section*{Acknowledgements}
This work was supported by the National Sciences and Engineering Research Council of Canada (NSERC) and Defence Research and Development Canada (DRDC).  We thank E. B. Ramirez for helpful discussions as well as S. Birner and Z. Wasilewski for assistance with software setup.

\appendix

\newpage
\section{Effect of electron occupancy on potential landscape }\label{app:schrodingerVSpoisson}

\begin{figure}[h]
    \centering
    \includegraphics[width = 0.5\textwidth]{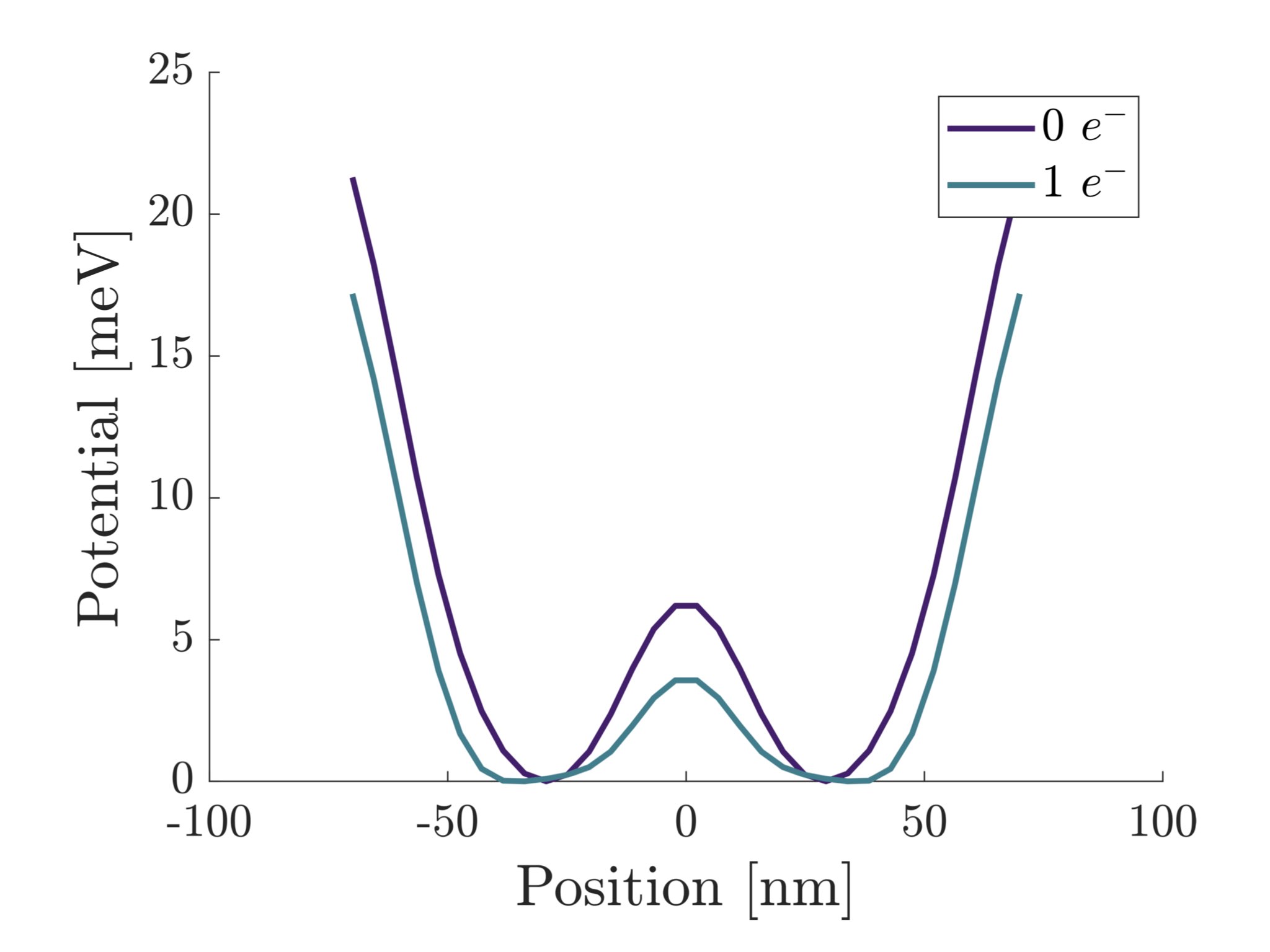}
    \caption{Effect of a single electron on a self-consistent calculation of a double well potential. The 1D 0 $e^-$ (purple) and 1 $e^-$ (blue) potentials were extracted from a 3D self-consistent Poisson calculation and a 3D self-consistent Schr\"{o}dinger-Poisson calculation respectively. The electric field from the single electron lowers the barrier height between the two potential wells which raises the inter-dot tunnel coupling while simultaneously widening the individual potential well widths which lower the QD orbital spacing.}
    \label{fig:effOfCharge}
\end{figure}

\indent In this paper, we used self-consistent 3D finite element calculations to map the set of applied gate voltages $\vec{V}$ to the electrostatic potential landscape of a QD device. The calculations were done in nextnano++ using either a Poisson (P) or Schr\"{o}dinger-Poisson (SP) solver.  SP calculations provide more realistic simulations of nanoscale QD structures as they properly model the quantum effect of the accumulated electron density on the potential landscape. \\
\indent Electron shuttling through a linear QD chain can be realized by sweeping the inter-dot detuning using the plunger gates that define each QD.  In a real linear QD chain, sweeping the plunger gates does not change the total electron occupancy if the QDs are well separated from a nearby electron reservoir. However, in a P or SP simulation, the electron density can change continuously as the plunger gate are swept. This varying electron density makes it difficult to compare potential landscapes for different plunger gate voltage configurations. Ideally, the integrated electron density could remain fixed during the SP calculation so the impact of the accumulated electron on the electrostatic potential is consistent; however, this is not possible in standard P and SP calculations.  For the shuttling simulations done in section~\ref{sec:coherentShuttling}, the plunger gate voltages were tuned below the device's turn-on voltage so that the integrated electron density was 0 $e^-$ for all voltage configurations. In this 0 $e^-$ regime, P and SP calculations provide the same potential landscape and is why P calculations were use in section~\ref{sec:coherentShuttling}.\\
\indent Next, we quantify how neglecting the electron impacts the electron shuttling results presented in the main text.  To do so, a 3D simulation of a double QD was done in nextnano++ and the respective plunger gate voltages were set equal to form a symmetric double well potential.  The plunger gate voltages were tuned to both the 0 $e^-$ regime just below the device turn-on and the 1 $e^-$ regime and are simulated using a P and SP calculation respectively.  1D extractions of the 3D potentials are plotted in Figure~\ref{fig:effOfCharge}. The 1D cuts are taken 1 nm below the Si/SiO$_{2}$ interface along the of QD chain (white dashed line in Figure~\ref{fig:QDchainSchematic}b).  The addition of the electron increases the inter-dot tunnel coupling from $t_c \approx 25$ $\mu$eV (0 $e^-$) to $t_c \approx 60$ $\mu$eV (1 $e^-$) due a reduction in the barrier height. Increasing $t_c$ does not qualitatively impact the results shown in section~\ref{sec:coherentShuttling}, and simply increases the shuttle speed by reducing the constant-adiabaticity pulse length.  The electron has the additional effect of widening the individual potential wells which decreases the orbital spacing for each QD.  For QD geometries where the orbital spacing is much greater than $t_c$, the lower orbital spacing will not affect shuttling performance.  However, if the orbital spacing and $t_c$ are comparable, then the lower orbital spacing can cause longer shuttling pulses in order to maintain adiabaticity as seen in section~\ref{sec:optimalGeom}. 

\clearpage
\section{Comparison of adiabatic and non-adiabatic orbital shuttling simulations}\label{app:simResults}

\begin{figure}[h]
    \centering
    \includegraphics[width = 0.93\textwidth]{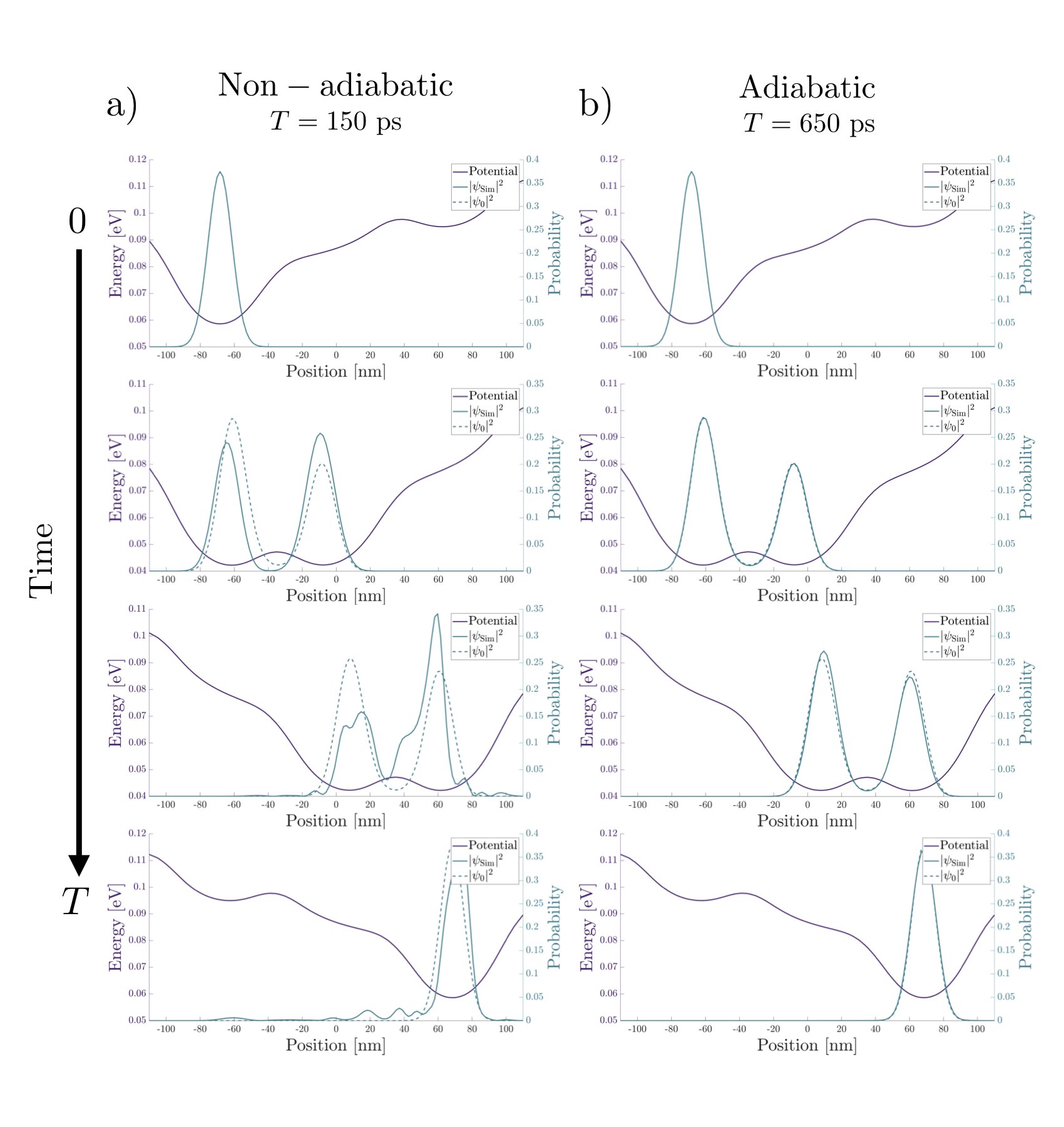}
    \caption{Electron shuttling simulations through a linear triple QD chain showing non-adiabatic (left column) and adiabatic (right column) evolution.  The non-adiabatic (adiabatic) pulse was calculated using $\xi = 0.08$ ($\xi = 0.02$) giving a total pulse length $T = 150$ ps ($T = 650$ ps). For both simulations, the electron is initialized initialized in the orbital ground state and simulated according to the time-dependent Shcr\"{o}dinger equation using the split-operator method with a time step $\Delta t = $ \num{5E-16} s. In each panel the instantaneous potential landscape (solid purple), orbital ground state (dashed blue) and simulated orbital state (solid blue) are plotted.  In the non-adiabatic case, the simulated orbital wave function deviates from the orbital ground state due to the faster pulse's sweep rate. Conversely, in the adiabatic case, the simulated orbital state remains in the orbital ground state throughout the entire pulse sequence.}
    \label{fig:appNAvsA}
\end{figure}
    
\newpage
\section{Orbital energy spacings from finite-element calculations}
\label{app:powerLawFit}
    
\begin{figure}[h]
    \centering
    \includegraphics[width = 0.5\textwidth]{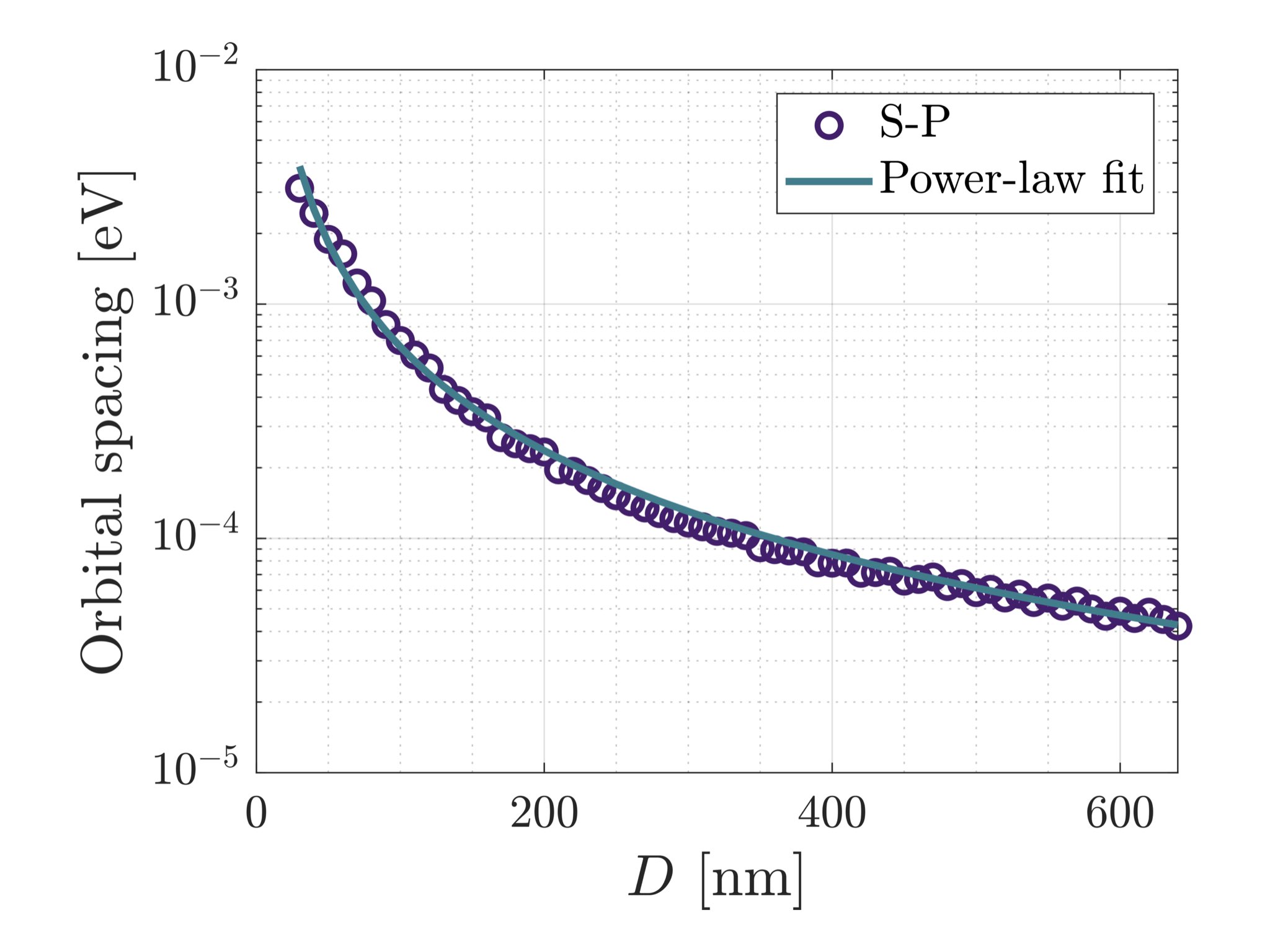}
    \caption{Orbital spacing versus plunger gate length $D$.  The orbital spacings are extracted using a self-consistent Schr\"{o}dinger-Poisson calculation and fit to a power-law $f(D) = aD^b$ with $a = 0.58$  ${\rm eV}\times {\rm nm}^{b}$ and $b = -1.47$.}
    \label{fig:appPowerLaw}
\end{figure}

Here, the orbital spacing as a function of plunger gate length $D$ is calculated for a simplified metal gate geometry QD. This is done by simulating a triple linear QD chain in nextnano++ using a finite element Schr\"{o}dinger-Poisson solver.  Thre three plunger gates have the same length $D$, a 40 nm width, and a 30 nm edge-to-edge separation.  The middle plunger gate is used to define the potential well and the outside plunger gates are included to model the impact of the QD chain.  The outer plunger gates are offset -0.1 V with respect to the middle plunger gate voltage $V$.  For a given plunger gate length $D$, $V$ is tuned so that the integrated electron density is $1\pm0.05$ $e^-$ in the QD.  From the resulting simulation, the energy difference between the ground and first excited energy states give the orbital spacing $\Delta E$. Figure \ref{fig:appPowerLaw} shows how $\Delta E$ varies with $D$.  The data are fit to a power-law $f(D) = aD^b$ with $a = 0.58$  ${\rm eV}\times {\rm nm}^{b}$ and $b = -1.47$ which is subsequently used in the optimal device geometry simulations in section~\ref{sec:optimalGeom}.

\newpage
\section{The spin-valley-orbit Hamiltonian}
\label{app:effSVOHam}
\begin{figure}[h]
    \centering
    \includegraphics[width = 0.6\textwidth]{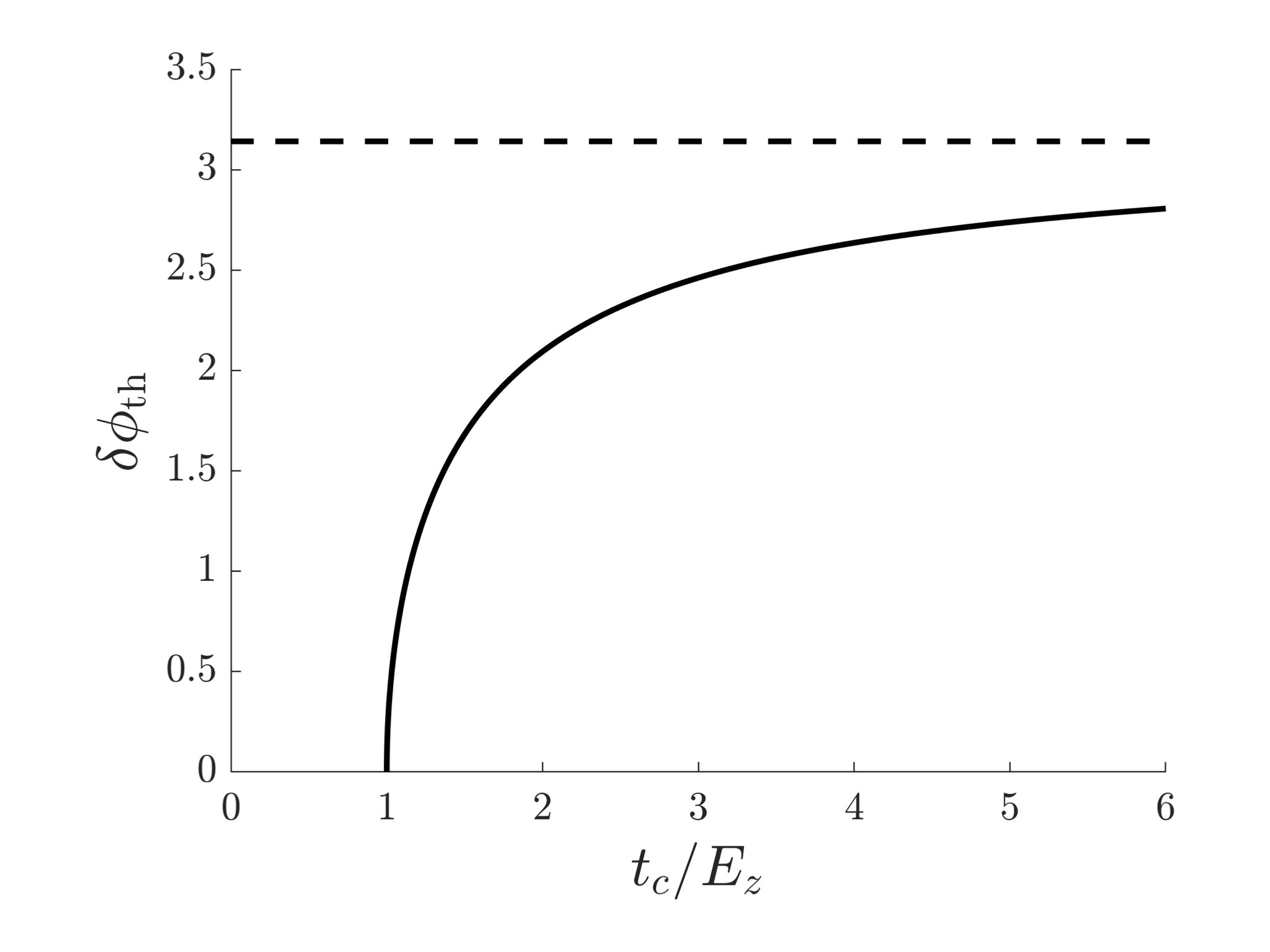}
    \caption{Relationship between the ratio $t_c/E_z$ and the threshold value of valley phase difference, $\delta\phi_{\rm th}$, where strong SVO mixing occurs.}
    \label{fig:deltaPhiThresh}
\end{figure}

Here we show how the valley phase difference affects the inter-dot tunnel coupling and spin-orbit interaction strengths in the effective Hamiltonian used in section~\ref{sec:effSimulations}. First consider a double quantum system consisting of two orbital states $\ket{L}$ and $\ket{R}$ where each state corresponds to the electron orbital occupying the left or right QD. Both QDs have a ground state energy $\epsilon_d$ (with $d = L,R$) and are coupled with strength $t_c$. Each QD has its own complex valley splitting $\Delta_d = |\Delta_d|e^{i\phi_d}$ which couples the two valley states $\ket{z}$ and $\ket{\tilde{z}}$.  The Hamiltonian of this valley-orbit subspace has the form
\begin{equation}
    H_{VO} = \sum_{d = L,R} \epsilon_d k_d \otimes \tau_0 + \sum_{d = L,R} (\Delta_d k_d \otimes \tau_+ + h.c) + t_c k_x \otimes \tau_0
\end{equation}
where the operators are all defined the same as in Eq.~\ref{eq:effHamiltonian} in the main text. The basis states are $\{d,\nu\}$ with $d = L,R$ (left and right orbital ground states) and $\nu = z, \tilde{z}$ (valley states).\\
\indent This Hamiltonian can be re-written into the valley eigen-basis $\ket{d,\pm} = \ket{z} \pm e^{i\phi_d}\ket{\tilde{z}}$ via the matrix transformation $B = \sum_{L,R} k_d \otimes B_d$ where
\begin{equation}
    B_d = \begin{bmatrix}
    1 & 1 \\
    e^{-i\phi_d} & -e^{-i\phi_d}
    \end{bmatrix}
\end{equation}
After the change of basis, the valley-orbit Hamiltonian becomes
\begin{multline}
    H_{VO}' = B^\dagger H_{VO}B = \sum_{d = L,R} \epsilon_d k_d \otimes \tau'_0 + \sum_{d = L,R} |\Delta_d| k_d \otimes \tau'_z \\
    + (t_{c,+}k_+ \otimes \tau'_0 + h.c) + (t_{c,-}k_- \otimes \tau'_x + h.c) 
\end{multline}
where $\tau'_i$ are two-level Pauli operators acting on this new valley eigen-basis. 
Setting $\phi_L = 0$ and $\phi_R = \delta\phi$ gives $t_{c,+} = \frac{t_c}{2}(1 + e^{-i\delta\phi})$ and $t_{c,-} = \frac{t_c}{2}(1 - e^{-i\delta\phi})$. 
The basis states are defined by $\{d,\nu'\}$ with $d = L,R$ and $\nu' = -, +$ (valley eigen-states). 
In this representation, the tunnel coupling $t_c$ term has transformed into two distinct forms: $t_{c,+}$ and $t_{c,-}$.
The $t_{c,+}$ and $t_{c,-}$ coupling terms couple orbital states only from the same and different valley eigen-states respectively. 
Their respective coupling strengths depend on the magnitude of the valley phase difference $\delta\phi$.
From the expression for $t_{c,+}$ ($t_{c,-}$), the coupling between orbital states from the same (different) valley eigen-state varies from being strongest (0) when $\delta\phi = 0$ to 0 (strongest) when $\delta\phi = \pi$.\\
\indent Next, the Hamiltonian is extended to include the electron's spin degree of freedom. This is the same Hamiltonian for the shuttled electron from Equation \ref{eq:effHamiltonian}. Applying the transformation $B$ onto this Hamiltonian results in
\begin{multline}
    \label{eq:effHamiltonian_valOrbBasis}
    H' = \sum_{d = L,R} \epsilon_d k_d \otimes \tau'_0 \otimes s_0 + \sum_{d = L,R} |\Delta_d| k_d \otimes \tau'_z \otimes s_0 + E_z k_0 \otimes \tau'_0 \otimes s_z \\
    + (t_{c,+} k_+ \otimes \tau'_0 \otimes s_0 + h.c) + (t_{c,-}k_- \otimes \tau'_x \otimes s_0 + h.c) \\
    + \eta_1 k_z \otimes \tau'_0 \otimes s_x + (\eta_{2,+} k_+ \otimes \tau'_0 \otimes is_y + h.c) + (\eta_{2,-} k_+ \otimes \tau'_x \otimes is_y + h.c)
\end{multline}
where $\eta_{2,+} = \frac{\eta_2}{2}(1 + e^{-i\delta\phi})$ and $\eta_{2,-} = \frac{\eta_2}{2}(1 - e^{-i\delta\phi})$. The basis states are now $\{d,\nu',s\}$ with $d = L,R$, $\nu' = -, +$ and $s = \uparrow, \downarrow$.\\
\indent When written in the valley eigen-basis, it is easier to see how the strength of the spin-orbit coupling terms $\eta_1$ and $\eta_2$ depend on the magnitude of the valley phase difference $\delta\phi$. The $\eta_2$ spin-orbit coupling term takes on two different forms $\eta_{2,+}$ and $\eta_{2,-}$, similarly to how the tunnel coupling $t_c$ transforms in this valley eigen-basis as well. These two terms $\eta_{2,+}$ and $\eta_{2,-}$ couple spin-orbit states from either the same or different valley eigen-state respectively. The coupling strengths of $\eta_{2,+}$ and $\eta_{2,-}$ are directly controlled by the magnitude of $\delta\phi$. When $\delta\phi = 0$, $\eta_{2,+}$ (which couples spin-orbital states from the same valley eigen-state) is maximal in strength. However as $\delta\phi \rightarrow \pi$, $\eta_{2,+}$ is suppressed and fully turned off at $\delta\phi = 0$. The converse is true for $\eta_{2,-}$ which couples spin-orbital states from different valley eigen-states. When $\delta\phi \rightarrow 0$, $\eta_{2,-}$ becomes suppressed, and $\eta_{2,-}$ is maximal when $\delta\phi = \pi$.\\
\indent For the $\eta_1$ term, when the detuning is large $|\epsilon| \gg 0$, the orbital eigen-states are $\ket{L}$ and $\ket{R}$. The $\eta_1$ term then acts as a single spin $X$ rotation operator with no dependence on $\delta\phi$. As the detuning is swept near the $t_{c,+}$ and $t_{c,-}$ anti-crossings, the orbital eigen-states become a super-position of $\ket{L}$ and $\ket{R}$, and the $\eta_1$ term couples both the orbital and spin states. The strength of the spin-orbit coupling near these anti-crossing is tied to the strength of $t_{c,+}$ and $t_{c,-}$. As $t_{c,+}$ increases (decreases), spin-orbital coupling between spin-orbital states from the same valley eigen-state is stronger (suppressed). Similarly as $t_{c,-}$ increases (decreases), spin-orbital coupling between spin-orbital state from different valley eigen-states is stronger (suppressed). Because both $t_{c,+}$ and $t_{c,-}$ depend on $\delta\phi$, the spin-orbit coupling from $\eta_1$ near these anti-crossing depends on $\delta\phi$ as well. \\
\indent When $t_c < E_z$, the ground orbital, excited spin state overlaps with the excited orbital, ground spin state, and the spin-orbit terms $\eta_{1,2}$ cause SVO mixing. When $t_c > E_z$, these states do not overlap until $\delta\phi$ reaches a threshold value $\delta\phi_{\rm th}$. For $\delta\phi < \delta\phi_{\rm th}$, the SVO mixing from $\eta_{1,2}$ is heavily suppressed. The threshold value $\delta\phi_{\rm th}$ occurs at $E_z = |t_{c,+}| = \frac{t_c}{2}|1+e^{i\delta\phi_{\rm th}}|$. The relationship between $t_c/E_z$ and $\delta\phi_{\rm th}$ is plotted in Figure \ref{fig:deltaPhiThresh}. A larger $t_c/E_z$ ratio provides greater tolerance for variations in $\delta\phi$ during shuttling. For $t_c \gg E_z$, $\delta\phi_{\rm th}$ approaches $\pi$.   

\newpage
\section{Validity of the single valley approximation in charge shuttling} \label{app:singleValleyApprox}
\begin{figure}[h]
    \centering
    \includegraphics[width = 0.6\textwidth]{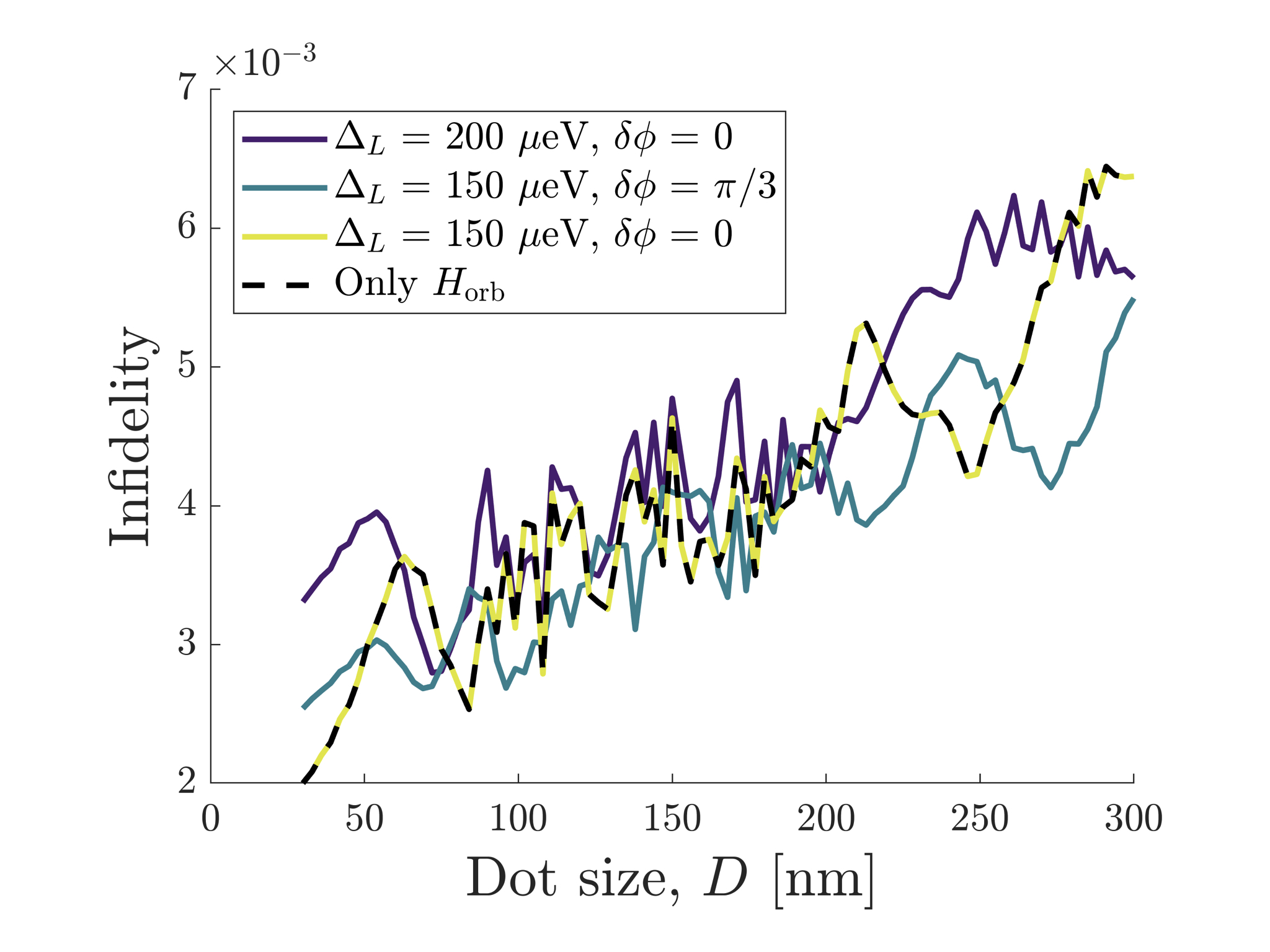}
    \caption{Fidelity of shuttling a charge with and without valley physics, versus dot size. The fixed Hamiltonian parameters are $t_c =$ 50 $\mu$eV and $|\Delta_R| =$ 150 $\mu$eV. The orbital state infidelity is plotted on the vertical axis, defined as $1-|{\rm Tr}[\rho_{\rm orb}(T)|R,0\rangle \langle R,0|]|^2$ where $\rho_{\rm orb}(T)$ is the orbital density matrix after shuttling. The dashed black line corresponds to simulations of $H_{\rm orb}$ where no valley physics is considered. Simulations of $H_{VO}$ which include valley physics are done for cases: equal valley splitting and no valley phase difference (yellow), non-equal valley splitting and no valley phase difference (purple), and equal valley splitting and non-zero valley phase difference (blue-green).}
    \label{fig:singValleyApprox}
\end{figure}

\indent Here we justify the the single-valley approximation used in the charge shuttling simulations in section~\ref{sec:coherentShuttling}.
First, consider the single-valley effective Hamiltonian used in Eq.~\ref{eq:effHamiltonian_excitedOrb} which includes both the ground and first excited state orbital degrees of freedom for a double quantum dot system.
\begin{equation}
    H_{\rm orb} = \begin{bmatrix} \epsilon_L & 0 & t_c & t_c \\
    0 & \epsilon_L + \Delta E_L & t_c & t_c \\
    t_c & t_c & \epsilon_R & 0 \\
    t_c & t_c & 0 & \epsilon_R + \Delta E_R
    \end{bmatrix}
\end{equation}
$\Delta E_d$ with $d = L,R$ is the orbital spacing in the left and right dots, and $t_c$ is the inter-dot tunnel coupling.
The basis states of $H_{\rm orb}$ are $\{d,n\}$ where $d = L,R$ (left and right QD) and $n = 0,1$ (ground and first excited state).
This effective Hamiltonian can be generalized to include valley states as
\begin{equation}
    H_{VO} = H_{\rm orb}\otimes\tau_0 + \sum_{d = L,R} (\Delta_d k_d \otimes I_2 \otimes \tau_+ + h.c.)
\end{equation}
where the $k$ operator acts on the left and right QD state subspace and $\tau$ acts on the valley subspace as described in Eq.~\ref{eq:effHamiltonian}.
The $2\times 2$ identity operator denoted $I_2$ acts on the ground and first excited state subspace.
The basis states of this Hamiltonian are $\{d,n,\nu\}$, where $\nu = z,\tilde{z}$ for the valley states.\\
\indent Constant-adiabaticity charge shuttling simulations are performed using both $H_{\rm orb}$ and $H_{VO}$ in order compare the dynamics resulting from the single-valley and the valley-orbit effective Hamiltonians.
To begin, a constant-adiabaticity pulse with $\xi = 0.005$ in which the detuning $\epsilon = \epsilon_L - \epsilon_R$ varies from -1.5 meV to 1.5 meV is calculated using $H_{\rm orb}$.
For both $H_{\rm orb}$ and $H_{VO}$, the state is initialized in the ground state of the respective Hamiltonian and subsequently evolved according to the pulse shape.
The infidelity of the shuttling process is calculated as $1-|{\rm Tr}[\rho_{\rm orb}(T)|R,0\rangle \langle R,0|]|^2$ where $\rho_{\rm orb}(T)$ is the orbital density matrix after shuttling.
For simulations of $H_{VO}$, $\rho_{\rm orb}(T)$ is found by tracing out the valley degree of freedom of the final density matrix.\\
\indent The fixed Hamiltonian parameters used are $t_c$ = 50 $\mu$eV and $|\Delta_R|$ = 150 $\mu$eV.
The dots are assumed to be of equal size, so $\Delta E_L = \Delta E_R$.
Dot size is mapped to an orbital energy spacing $\Delta E$ using the fit parameters from Appendix~\ref{app:powerLawFit}.
Figure~\ref{fig:singValleyApprox} shows the infidelity of the shuttling simulations as a function of dot size for different combinations of $\delta\phi$ and $|\Delta_L|$.
The infidelity increases slightly with dot size in all cases; this is due to increased overlap of the final state with the excited orbital states $\ket{L,1}$ and $\ket{R,1}$ with increasing dot size.
We see that the single-valley Hamiltonian (dashed black) and the valley Hamiltonian produce the same shuttling dynamics when $\Delta_L =$ $\Delta_R =$ 150 $\mu$eV and $\delta\phi = 0$ (solid yellow).
When $\Delta_L \neq \Delta_R$ (solid purple) or $\delta\phi \neq 0$ (solid blue-green), the two Hamiltonians produce different dynamics, but with the infidelities of similar order.
This confirms that the charge shuttling simulations presented in section~\ref{sec:coherentShuttling}, which only took into account the single-valley orbital states, should give a reasonable approximation to the physically relevant case of charge shuttling in the presence of valley states.
However this is only strictly true if the interdot valley splittings are equal and there is no valley phase difference.

\clearpage
\section{Effective Hamiltonian simulations with $t_c < E_z$} \label{app:effResultsContinued}    
\begin{figure}[h]
    \centering
    \includegraphics[width = \textwidth]{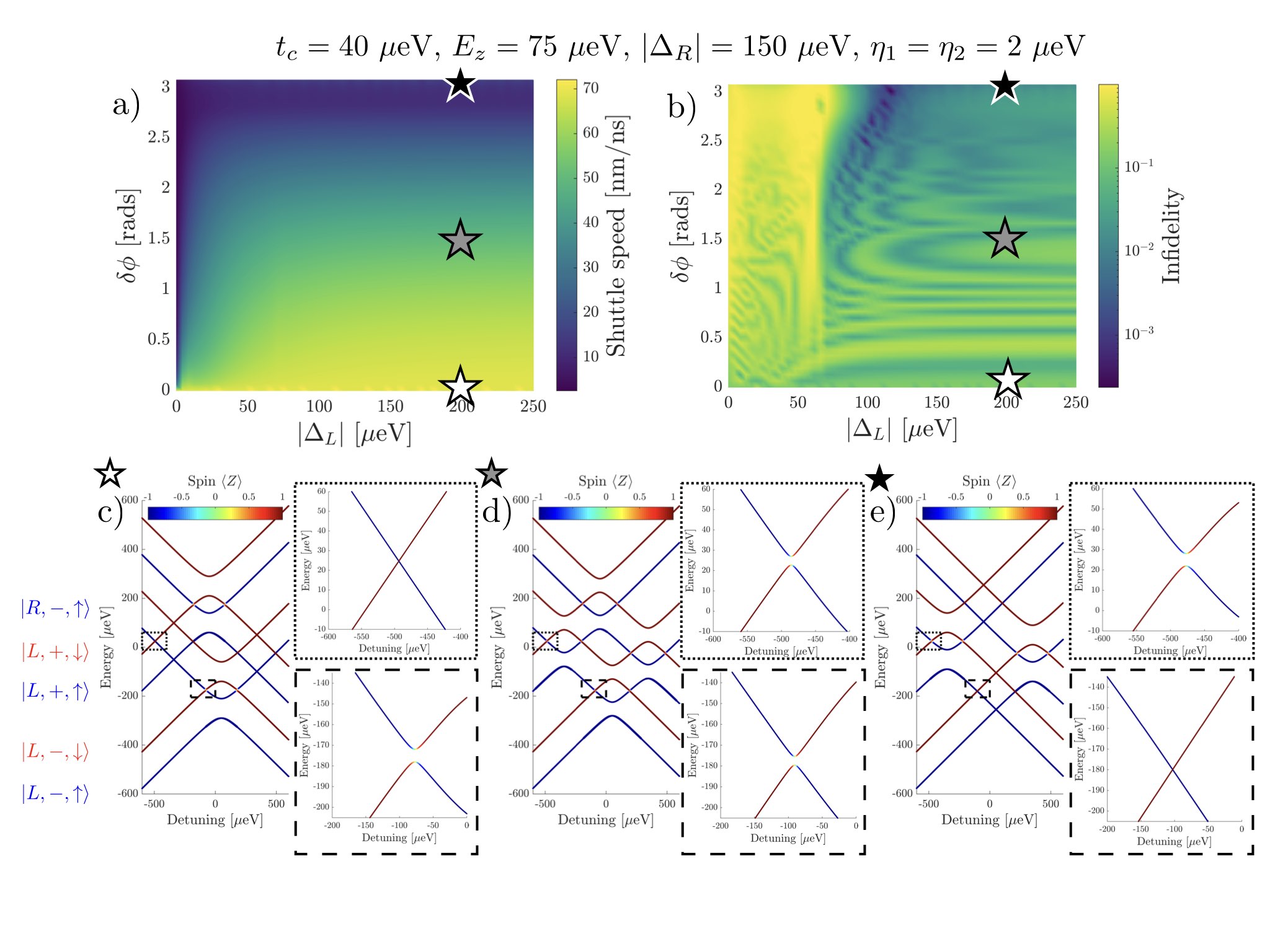}
    \caption{Shuttling one member of a singlet pair, for $t_c < E_z$.  For all panels, the fixed Hamiltonian parameters are: $t_c$ = 40 $\mu$eV, $E_z$ = 75 $\mu$eV, $|\Delta_R|$ = 150 $\mu$eV, $S_1$ = $S_2$ = 2 $\mu$eV. a) Variation of shuttle speed with the left QD valley splitting $|\Delta_L|$ and the inter-dot valley phase difference $\delta\phi$. b) Final singlet state fidelity's dependence on $|\Delta_L|$ and $\delta\phi$. The infidelity is plotted in colour scale, defined as $1 - |{\rm Tr}[\rho(T)(I_4\otimes\ket{S}\bra{S})]|^2$, where $\rho(T)$ is the density matrix after shuttling and $I_4$ is the $4\times4$ identity matrix. Dark blue indicates high fidelity shuttling, yellow indicates low fidelity.  c), d) and e) show the shuttled electron energy spectrum versus detuning. $|\Delta_L| = 200$ $\mu$eV in all panels, and $\delta\phi =$ $0$, $\frac{\pi}{2}$, and $\pi$ for (c), (d), and (e) respectively.  Energy levels are labelled according to their initial energy eigen-state when the detuning $\ll 0$. Enlarged views near the $t_{c,+}$ (dashed square) and $t_{c,-}$ (dotted square) anti-crossings demonstrate how the SVO mixing varies with $\delta\phi$.}
    \label{fig:effResultstcLTEz}
\end{figure}

\indent Here we extend the results from section~\ref{sec:effSimulations} and present electron singlet shuttling simulations using an effective Hamiltonian where $t_c < E_z$. The fixed Hamiltonian parameters are $t_c$ = 40 $\mu$eV, $E_z$ = 75 $\mu$eV, $|\Delta_R|$ = 150 $\mu$eV, and $\eta_1$ = $\eta_2$ = 2 $\mu$eV. The electron pair is initialized into the state $\ket{\psi(0)} = \frac{1}{\sqrt{2}}\ket{\psi^{VO}_0(0)}\otimes(\ket{\uparrow\downarrow} - \ket{\downarrow\uparrow})$ where $\ket{\psi^{VO}_0(0)}$ is the ground state of the initial valley-orbit Hamiltonian, and simulation is via the time time-dependent Schr\"{o}dinger equation. Figure~\ref{fig:effResultstcLTEz} shows how the shuttle speed and post-shuttling singlet state fidelity vary with $\delta\phi$ and $|\Delta_L|$. Both the transport speed and final spin state infidelity are calculated as described in section~\ref{sec:effSimulations}.\\
\indent Figure \ref{fig:effResultstcLTEz}a shows how shuttle speed varies with $|\Delta_L|$ and $\delta\phi$. $|\Delta_L|$ is varied from 0.1 - 250 $\mu$eV and $\delta\phi$ is varied from $[0,\pi)$.  Here the shuttle speed decreases and $\delta\phi$ increases similarly to the case when $t_c > E_z$ which is discussed in section~\ref{sec:effSimulations} and is caused by the closing $t_{c,+}$ anti-crossing.  Figures \ref{fig:effResultstcLTEz}c-e show the shuttled electron energy spectra versus detuning at $|\Delta_L| = 200$ $\mu$eV and $\delta\phi =$ $0$, $\frac{\pi}{2}$, and $\pi$ respectively. The initially $\ket{L,-\uparrow}$ and $\ket{R,-,\uparrow}$ energy levels form the $t_{c,+}$ anti-crossing near zero detuning which closes as $\delta\phi$ increases. When $|\Delta_L| > |t_{c,+}|$, the valley splitting $|\Delta_L|$ has little impact on the shuttle speed. This is because the duration of the constant-adiabaticity pulse is determined by the smallest energy scale with respect to the ground state which in this regime is the $t_{c,+}$ anti-crossing. However, when $|\Delta_L| < |t_{c,+}|$, the valley splitting becomes the smallest energy scale with respect to the ground state and controls the shuttle speed. The overall shuttle speed is lower here for $t_c < E_z$ compared to the $t_c > E_z$ simulations in section~\ref{sec:effSimulations} because $t_c$ is smaller ($40$ $\mu$eV compared to $75$ $\mu$eV).\\
\indent Figure \ref{fig:effResultstcLTEz}b shows the final singlet state fidelity dependence on $|\Delta_L|$ and $\delta\phi$. The colour scale corresponds to the final singlet state infidelity.  Unlike when $t_c > E_z$ in section~\ref{sec:effSimulations}, there is no clear region of good fidelity singlet shuttling.  Here where $t_c < E_z$, the first and second excited energy levels cross as the detuning is swept for all values of $\delta\phi$ as seen in Figures~\ref{fig:effResultstcLTEz}c-e. These crossing cause SVO mixing from the $\eta_1$ and $\eta_2$ Hamiltonian terms occurs which reduces the singlet fidelity. In the region where $|\Delta_L| > E_z$, the singlet fidelity does improve modestly with $\delta\phi$. This is caused by a reduction of the SVO mixing strength between the same valley eigenstates as $\delta\phi$ increases (refer to Appendix \ref{app:effSVOHam} for details).  While the singlet fidelity does improve at higher $\delta\phi$, it is an undesirable region for electron shuttling as both the shuttle speed is reduced and there is non-zero spin-orbit mixing until $\delta\phi = \pi$.\\
\indent When $|\Delta_L| < E_z$, the final singlet state fidelity is low for all $\delta\phi$. In this regime, the shuttled electron's energy spectra becomes more complicated compared to the spectra shown in Figure \ref{fig:effResultstcLTEz}c-e. Here, the initially $\ket{L,-,\downarrow}$ energy level always crosses with spin-orbit energy levels from both the same valley eigen-state and different valley eigen-states. Because the electron travels along the $\ket{L,-,\downarrow}$ energy level during shuttling, it experiences both types of SVO mixing (between the same and different valley eigenstates). As $\delta\phi$ increases, the SVO mixing occurring between the same valley eigenstate decreases and is strongest when $\delta\phi = 0$.  Conversely, as $\delta\phi$ decreases, the SVO mixing between different valley eigenstates increases and is strongest when $\delta\phi = \pi$. Because the electron sees both types of SVO mixing during shuttling, there is never a value of $\delta\phi$ where the SVO mixing is suppressed to give good singlet shuttling fidelity.

\newpage
\section{Effective Hamiltonian simulations with $t_c > E_z$ and weaker spin-orbit coupling} \label{app:effResultsContinuedSmallSO}  
\begin{figure}[h]
    \centering
    \includegraphics[width = \textwidth,trim={0 25cm 0 0},clip]{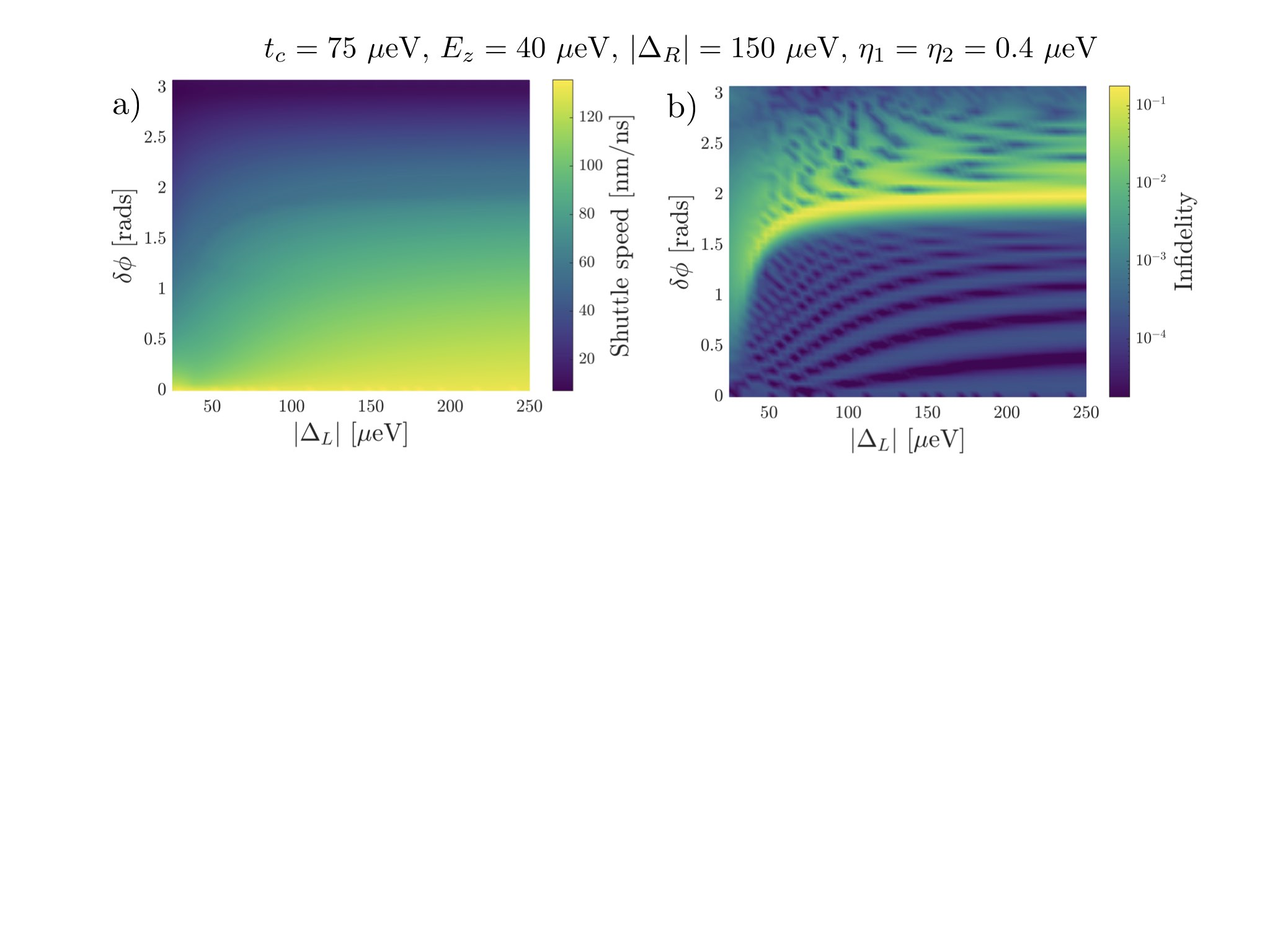}
    \caption{Shuttling one member of a singlet pair, for $t_c < E_z$.  For all panels, the fixed Hamiltonian parameters are: $t_c$ = 40 $\mu$eV, $E_z$ = 75 $\mu$eV, $|\Delta_R|$ = 150 $\mu$eV, $\eta_1 = \eta_2$ = 0.4 $\mu$eV. a) Variation of shuttle speed with the left QD valley splitting $|\Delta_L|$ and the inter-dot valley phase difference $\delta\phi$. b) Final singlet state fidelity's dependence on $|\Delta_L|$ and $\delta\phi$. The infidelity is plotted in colour scale, defined as $1 - |{\rm Tr}[\rho(T)(I_4\otimes\ket{S}\bra{S})]|^22$, where $\rho(T)$ is the density matrix after shuttling and $I_4$ is the $4\times4$ identity matrix. Dark blue indicates high fidelity shuttling, yellow indicates low fidelity.}
    \label{fig:effResultstcGTEzSmallSO}
\end{figure}

\begin{figure}[h]
    \centering
    \includegraphics[width = 0.6\textwidth]{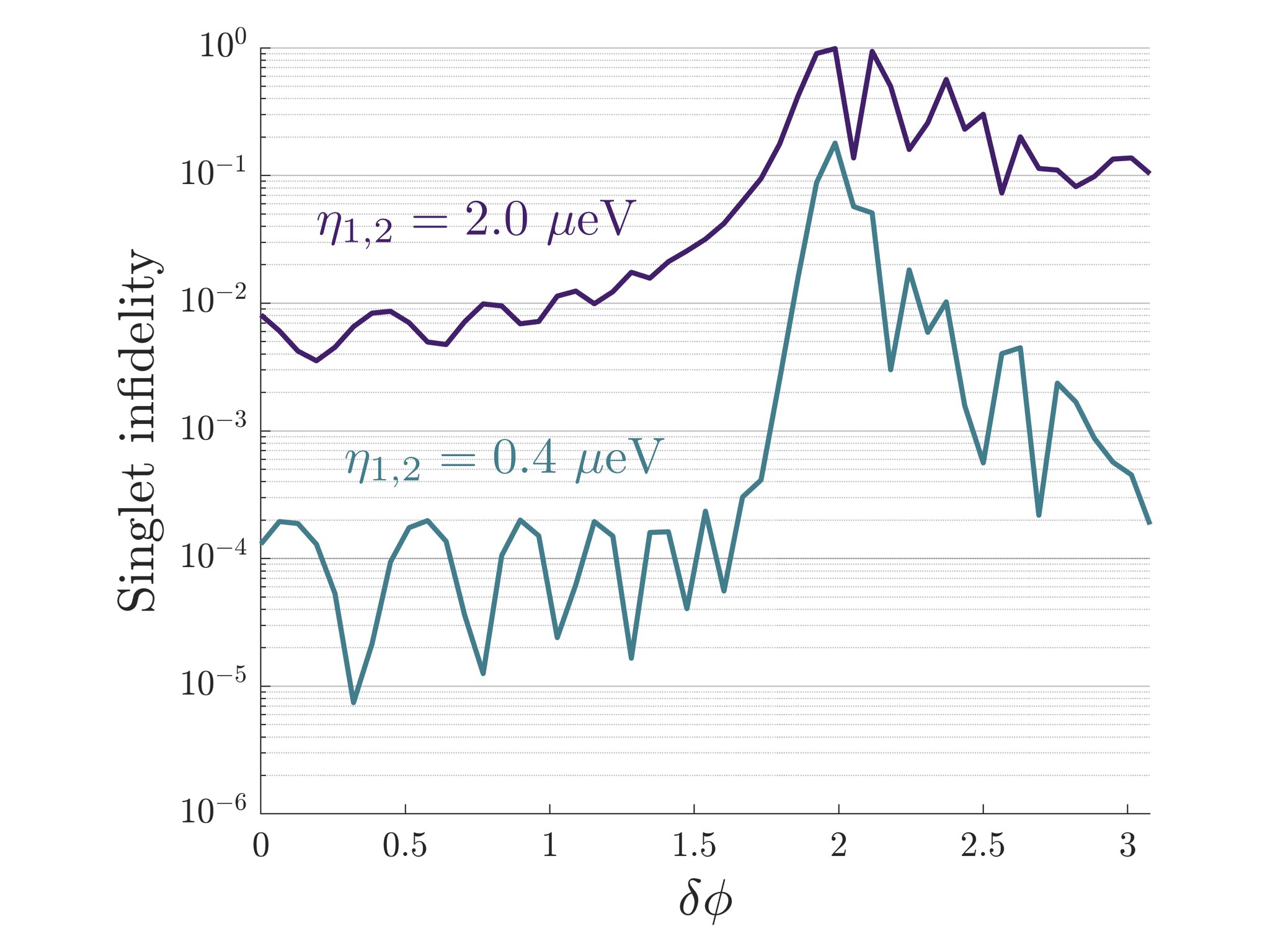}
    \caption{Comparison of shuttling singlet infidelity for two different spin-orbit coupling strengths: $\eta_{1,2} = 2\,\,\mu{\rm eV}$ (purple) and $\eta_{1,2} = 0.4\,\,\mu{\rm eV}$ (blue).  Traces are taken directly from Figures \ref{fig:effResultstcGTEz}b and \ref{fig:effResultstcGTEzSmallSO}b. These are the raw infidelities, with no single-spin correction rotations applied.}
    \label{fig:effResultstcGTEzSmallSOdirComp}
\end{figure}

\indent Results for constant-adiabaticity shuttling simulations where $\xi = 0.005$ similar to that described in section~\ref{sec:effSimulations} but with a smaller spin-orbit coupling $\eta_1 = \eta_2 = 0.4$ $\mu$eV are shown in Figure \ref{fig:effResultstcGTEzSmallSO}. The other static Hamiltonian parameters are $t_c$ = 40 $\mu$eV, $E_z$ = 75 $\mu$eV, and $|\Delta_R|$ = 150 $\mu$eV. The varied parameters are $\delta\phi \in [0,\pi)$ and $|\Delta_L| \in [25,250]$ $\mu$eV. The initial electron pair state, shuttle speed (a), and singlet infidelity (b) are calculated the same way as in section~\ref{sec:effSimulations}. Figure \ref{fig:effResultstcGTEzSmallSOdirComp} compares the singlet fidelity for the larger (purple, $\eta_{1,2} = 2\,\,\mu{\rm eV}$) and smaller (blue, $\eta_{1,2} = 0.4\,\,\mu{\rm eV}$) spin-orbit coupling simulations. The traces are taken from Figures \ref{fig:effResultstcGTEz}b (larger spin-orbit) and \ref{fig:effResultstcGTEzSmallSOdirComp} (smaller spin-orbit) along $\delta\phi$ at $|\Delta_L| = 200\,\,\mu{\rm eV}$. The smaller spin orbit simulation shows $\sim$1.5 orders of magnitude improvement in singlet fidelity for a factor of 5 reduction in spin-orbit strength.  

\end{document}